%% file: main.tex
\documentclass[aps,amsmath,amssymb,preprintnumbers,groupedaddress,twocolumn]{revtex4}
\usepackage{graphicx,psfrag,hyperref}
\usepackage{amsmath,bbm,array,amsfonts,graphicx,wrapfig,lscape,float,mathtools,multirow,longtable,amssymb}
\usepackage[dvipsnames]{xcolor}

\input{pref}

\begin{document}

\preprint{UNIST-MTH-24-RS-06}
\title{Generative AI for Brane Configurations and Coamoeba}

\author{Rak-Kyeong Seong}

\email{seong@unist.ac.kr}

\affiliation{\it
Department of Mathematical Sciences, and 
Department of Physics,\\ 
Ulsan National Institute of Science and Technology,\\
50 UNIST-gil, Ulsan 44919, South Korea
}

\begin{abstract}
We introduce a generative AI model to obtain Type IIB brane configurations that realize toric phases of a family of $4d$ $\mathcal{N}=1$ supersymmetric gauge theories. 
These $4d$ $\mathcal{N}=1$ quiver gauge theories
are worldvolume theories of a D3-brane probing a toric Calabi-Yau 3-fold. 
The Type IIB brane configurations are given by the coamoeba projection of the mirror curve associated with the toric Calabi-Yau 3-fold.
The shape of the mirror curve and its coamoeba projection, 
as well as the corresponding Type IIB brane configuration and the toric phase of the $4d$ $\mathcal{N}=1$ theory,
all depend on the complex structure moduli parameterizing the mirror curve. 
We train a generative AI model, a conditional variational autoencoder (CVAE), 
that takes a choice of complex structure moduli as input and generates the corresponding coamoeba.
This enables us not only to obtain a high-resolution representation of the entire phase space for a family of $4d$ $\mathcal{N}=1$ theories corresponding to the same toric Calabi-Yau 3-fold, but also to continuously track the movements of the mirror curve and the branes wrapping the curve in the corresponding Type IIB brane configurations during phase transitions associated with Seiberg duality.
\end{abstract} 
\maketitle
\noindent

\section{Introduction}

Since the pioneering works in \cite{He:2017aed, Krefl:2017yox, Ruehle:2017mzq, Carifio:2017bov, He:2017set}, 
machine learning has pushed the boundaries of what is computationally possible in quantum field theory and string theory \cite{Hashimoto:2018ftp, Cole:2018emh, Halverson:2019tkf, Hashimoto:2019bih, Cole:2019enn, Bies:2020gvf, Ruehle:2020jrk, Bao:2020nbi, Halverson:2020trp, Akutagawa:2020yeo, Cole:2021nnt,  Abel:2021rrj, Maiti:2021fpy, Seong:2023njx, Halverson:2024axc, Gukov:2024opc}
as well as in geometry and topology \cite{Bull:2018uow, Jejjala:2019kio, Brodie:2019dfx, He:2020lbz, Gukov:2020qaj, Anderson:2020hux, Berman:2021mcw, Ashmore:2021ohf, Berglund:2021ztg, Larfors:2021pbb, Berglund:2022gvm, Berglund:2023ztk, Gukov:2023kvx, Halverson:2023ndu, Choi:2023rqg}. 
In this work, we advance this development by introducing a generative AI model in order to study a family of Type IIB brane configurations that realize $4d$ $\mathcal{N}=1$ supersymmetric gauge theories corresponding to a toric Calabi-Yau 3-fold.
These $4d$ $\mathcal{N}=1$ theories are worldvolume theories of a D3-brane probing the toric Calabi-Yau 3-fold \cite{fulton1993introduction, Leung:1997tw}.
The Type IIB brane configurations that realize this family of $4d$ $\mathcal{N}=1$ theories are associated to bipartite periodic graphs on a 2-torus $T^2$, which have been studied in mathematics as \textit{dimer models} \cite{Kenyon:2003eyz, kasteleyn1967graph} and are known in string theory as \textit{brane tilings} \cite{Franco:2005rj, Hanany:2005ve, Franco:2005sm}.

\begin{figure*}[ht!!]
\begin{center}
\resizebox{0.95\hsize}{!}{
\includegraphics[height=5cm]{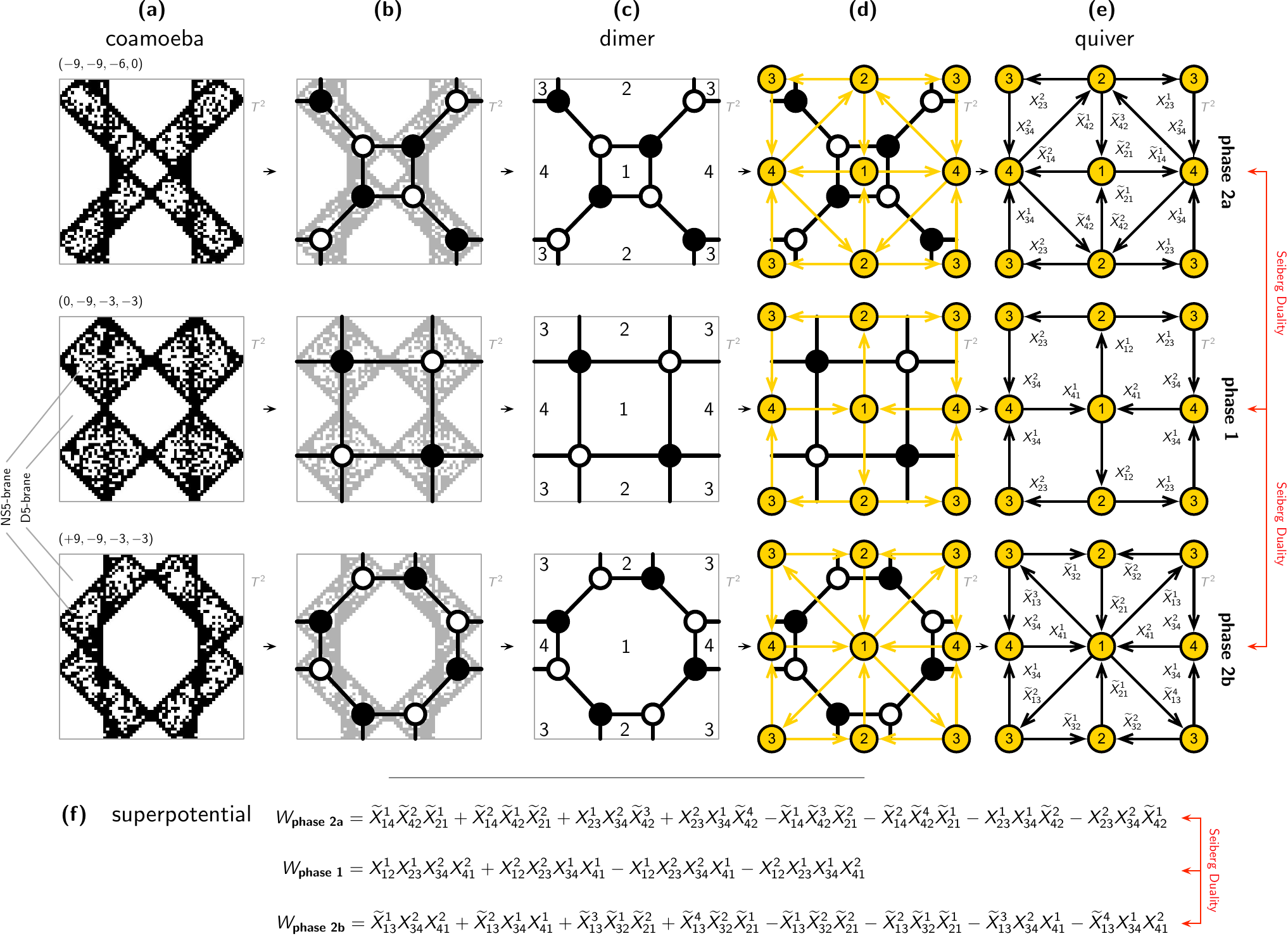} 
}
\caption{
(a) The coamoeba for the cone over the zeroth Hirzebruch surface $F_0$ are shown at 3 different values of the complex structure moduli $(c_{11}, c_{12}, c_{21}, c_{22})$. The white regions are occupied by the D5-brane, whereas the black regions are occupied by the NS5-brane wrapping $\Sigma$. 
(b)-(c) The skeleton graph of the coamoeba is a bipartite periodic graph on $T^2$ known as a dimer model or a brane tiling. This work concentrates on the coamoeba projection itself and its role in locating the D5- and NS5-branes in the Type IIB brane configuration as shown in (a).
In (c), the polygonal faces in the bipartite graph are labelled by integers $i=1, \dots, G$, where $G$ is the number of $U(N)_i$ gauge groups in the corresponding $4d$ $\mathcal{N}=1$ gauge theory. These faces are due to the bipartite nature of the graph even-sided and have therefore even number of boundary edges and nodes.
Each face corresponds to a $U(N)_i$ gauge group of the $4d$ $\mathcal{N}=1$ theory.
Edges connecting white and black nodes of the dimer correspond to bifundamental chiral fields $X_{ij}$, where the orientation around white nodes is clockwise and around black nodes is counter-clockwise. These orientations also give the correct gauge-invariant combinations of chiral fields around white and black nodes that are associated respectively to positive and negative terms in the superpotential $W$ of the $4d$ $\mathcal{N}=1$ theory.
(d)-(e) The dual graph of the brane tiling forms what we refer to as a periodic quiver on $T^2$.
The associated superpotentials are shown in (f). We note here that the continuous change of complex structure moduli changes the shape of the coamoeba on $T^2$, which in turn modifies the associated brane tilings and corresponding $4d$ $\mathcal{N}=1$ theories. These $4d$ $\mathcal{N}=1$ theories corresponding to the same toric Calabi-Yau 3-fold, here the cone over $F_0$, are related by Seiberg duality and represent toric phases. 
\label{f_fig00}}
 \end{center}
 \end{figure*}

As illustrated in \fref{f_fig00}, brane tilings on $T^2$ are bipartite skeleton graphs of what is known as the \textit{coamoeba projection} \cite{Kenyon:2003uj, Feng:2005gw} 
of the mirror curve $\Sigma$ onto $T^2$.
The mirror curve $\Sigma$ is associated with the toric Calabi-Yau 3-fold
and is parameterized in terms of \textit{complex structure moduli} \cite{Hori:2000ck, Hori:2000kt, horimirror}.
It is a holomorphic curve and 
plays a quintessential role in the Type IIB brane configuration
realizing the family of $4d$ $\mathcal{N}=1$ theories that we are studying in this work.
Under T-duality, the probe D3-brane at the Calabi-Yau 3-fold singularity
becomes a D5-brane suspended between a NS5-brane 
that wraps the holomorphic curve $\Sigma$ in the Type IIB brane configuration. 
Changes to the complex structure moduli in $\Sigma$ lead to
changes in the configuration of D5- and NS5-branes 
giving rise to transitions between distinct Type IIB brane configurations.
These transitions can be interpreted as \textit{Seiberg duality} \cite{Seiberg:1994pq} between corresponding \textit{toric phases} of the $4d$ $\mathcal{N}=1$ theories \cite{Feng:2000mi, Feng:2001bn, Feng:2001xr, Feng:2002zw}, as illustrated in \fref{f_fig00}. 

In \cite{Seong:2023njx}, 
we introduced unsupervised machine learning techniques such as \textit{principal component analysis (PCA)} \cite{pearson1901liii, hotelling1933analysis, jackson2005user, jolliffe2002principal, jolliffe1990principal, deisenroth2020mathematics} and \textit{$t$-distributed stochastic neighbor embedding ($t$-SNE)} \cite{hinton2002stochastic, van2008visualizing}
in order to obtain a representation for
the phase space for
$4d$ $\mathcal{N}=1$ supersymmetric gauge theories corresponding to the same toric Calabi-Yau 3-fold.
This was done by dimensionally reducing a set of coamoeba at different values of the complex structure moduli in $\Sigma$.
In the resulting phase diagram, 
toric phases of $4d$ $\mathcal{N}=1$ supersymmetric gauge theories
were depicted as clusters of points, where each point was corresponding to a coamoeba at a different choice of complex structure moduli in $\Sigma$.
Moreover, transitions between
clustered regions in the phase diagram were identified with Seiberg duality between different toric phases corresponding to the same toric Calabi-Yau 3-fold.

Motivated by the results in \cite{Seong:2023njx}, we introduce in this work a \textit{generative AI} model that is able to identify and produce accurately the coamoeba on $T^2$.
The generated coamoeba are parameterized in terms of complex structure moduli of the Calabi-Yau mirror curve $\Sigma$. 
The Type IIB brane configurations defined in terms of the coamoeba
realize $4d$ $\mathcal{N}=1$ supersymmetric gauge theories corresponding to the toric Calabi-Yau 3-folds. 
In this work, we demonstrate that we can train a generative AI model such that it learns a `world model' of all possible coamoeba for any values of the complex structure moduli in $\Sigma$.
It is important to note here that the results of this paper do not elaborate on a practical method for determining the corresponding brane tiling 
and $4d$ $\mathcal{N}=1$ quiver gauge theory directly from the coamoeba projection of $\Sigma$. 
This remains an interesting open problem that deserves further exploration in the future. 

The generative AI model that we introduce in this work is a \textit{conditional variational autoencoder (CVAE)} \cite{kingma2013auto, sohn2015learning, cinelli2021variational}, 
which is lighter and more importantly more interpretable and explainable than other generative models such as for example generative adversarial networks (GANs) \cite{goodfellow2014generative, Erbin:2018csv, Halverson:2020opj}, diffusion models \cite{sohl2015deep} and other transformer-based architectures \cite{vaswani2017attention, Gukov:2020qaj}. 
For instance, the \textit{latent space} of a CVAE is structured and given in terms of a continuous probability distribution
that allows direct access to latent space vectors in the latent space.
These latent space vectors can be dimensionally reduced to produce interpretable low-dimensional latent space representations. 
In this work, we provide an example of such a latent space representation, 
and give possible interpretations for the structure of the represented latent space in relation to the generated coamoeba and the associated Type IIB brane configurations.

We particularly focus in this work on the cone over the zeroth Hirzebruch surface $F_0$, which is a toric Calabi-Yau 3-fold.
By training the CVAE model on samples of coamoeba corresponding to $F_0$, 
we show that the CVAE model
is capable to generate accurately the coamoeba for any choice of complex structure moduli in $\Sigma$.
This allows us 
not only to obtain a near-continuous version of the phase diagram introduced in \cite{Seong:2023njx} for $F_0$, 
but also introduce trajectories within the phase space of $F_0$
that correspond to continuous deformations of the coamoeba and
continuous movements of branes in the associated Type IIB brane configurations. 
\\

\section{Type IIB Brane Configurations and $4d$ $\mathcal{N}=1$ Quiver Gauge Theories}

\begin{table}[H]
\begin{center}
\begin{tabular}{|c|cccc|cccc|cc|}
\hline
\; & 0 & 1 & 2 & 3 & 4 & 5 & 6 & 7 & 8 & 9 \\
\hline
\text{D5} & $\times$ & $\times$ & $\times$ & $\times$ 
& $\times$ & $\cdot$ & $\times$ & $\cdot$
& $\cdot$ & $\cdot$
\\
\text{NS5} & $\times$ & $\times$ & $\times$ & $\times$ 
& \multicolumn{4}{c|}{---$\Sigma$---}
& $\cdot$ & $\cdot$
\\
\hline
\end{tabular}
\caption{
The Type IIB brane configuration consisting of a D5-brane suspended between a NS5-brane wrapping a holomorphic curve $\Sigma$, which realizes a $4d$ $\mathcal{N}=1$ supersymmetric gauge theory corresponding to a toric Calabi-Yau 3-fold. 
\label{t_tab01}
}
\end{center}
\end{table}

In this work, we consider a family of $4d$ $\mathcal{N}=1$ supersymmetric gauge theories which are 
worldvolume theories of a D3-brane probing a toric Calabi-Yau 3-fold \cite{fulton1993introduction, Leung:1997tw}. 
These $4d$ $\mathcal{N}=1$ gauge theories are also realized by a Type IIB brane configuration
which is related to the D3-brane at the Calabi-Yau singularity by T-duality.
When one takes T-duality
along the $(46)$ directions,
the probe D3-brane becomes a D5-brane
which
extends along the $(012346)$ directions.
It is suspended from a NS5-brane that extends along the $(0123)$ directions and wraps a holomorphic curve $\Sigma$ embedded into the $(4567)$ directions as summarized in \tref{t_tab01}.

The holomorphic curve, also referred to as the \textit{mirror curve} \cite{Hori:2000ck, Hori:2000kt, horimirror}, is given by
\beal{es01a00}
\Sigma : P(x,y) = 0 ~,~
\eea 
where $P(x,y)$ is the \text{Newton polynomial} in $x,y \in \mathbb{C}^*$ corresponding to the toric diagram $\Delta$ of the probed toric Calabi-Yau 3-fold.
It is defined as, 
\beal{es01a01}
P(x,y) = \sum_{(n_x, n_y) \in \Delta} c_{(n_x, n_y)} x^{n_x} y^{n_y} ~,~
\eea
where $(n_x, n_y) \in \mathbb{Z}^2$ are the coordinates of the vertices of the toric diagram $\Delta$.
We refer to the complex coefficients $c_{(n_x, n_y)}$ as the \textit{complex structure moduli} of the mirror Calabi-Yau 3-fold defined through $P(x,y)$ \cite{Hori:2000ck, Hori:2000kt, horimirror}.
The complex structure moduli can be scaled such that overall 
only $|\Delta| - 3$ remain independent \cite{Franco:2016qxh, Franco:2016tcm}.
These independent complex structure moduli can be used to define a complex structure moduli vector of the form,
\beal{es01a01b}
\mathbf{c} = (c_1, \dots, c_{|\Delta|-3}) \in (\mathbb{C^*})^{|\Delta| -3}
~.~
\eea
The complex coordinates $x,y \in \mathbb{C}^*$ in the Newton polynomial $P(x,y)$ can be associated with the $(45)$ and $(67)$ directions, respectively.
In fact, the arguments $(\arg(x), \arg(y)) \in T^2$ are the $(46)$ directions
and are periodically identified to form a 2-torus $T^2$.

The holomorphic curve $\Sigma$ and the associated Type IIB brane configuration can be visualized when one projects $\Sigma$ onto $(\arg(x), \arg(y))=(\theta_x, \theta_y) \in T^2$, 
\beal{es01a02}
\Sigma &\rightarrow &
T^2
\nn\\
(x,y)=(r_x e^{i\theta_x}, r_y e^{i\theta_y}) &\mapsto&
(\theta_x, \theta_y) ~.~
\eea
This is also known as the \textit{coamoeba projection} of $\Sigma$ \cite{Kenyon:2003uj, Feng:2005gw} in tropical geometry. 
The locations on $T^2$ where the D5-brane meets the NS5-brane wrapping $\Sigma$ is visualized in the coamoeba projection of $\Sigma$. 
As shown in \fref{f_fig00}, the skeleton graph of the coamoeba on $T^2$ can be identified with a bipartite periodic graph on $T^2$ also known as a \textit{dimer} \cite{Kenyon:2003eyz, kasteleyn1967graph} or a brane tiling \cite{Franco:2005rj, Hanany:2005ve, Franco:2005sm}. 
There is a one-to-one correspondence between a coamoeba on $T^2$ and a brane tiling. The reader is referred to \cite{Feng:2005gw, Franco:2016qxh} for more details on this correspondence.

As illustrated in \fref{f_fig00}, 
the brane tiling associated with the coamoeba on $T^2$ encodes the quiver and superpotential of the $4d$ $\mathcal{N}=1$ supersymmetric gauge theory corresponding to the toric Calabi-Yau 3-fold.
Given that we have $N$ D3-branes probing the toric Calabi-Yau 3-fold, 
the even-sided faces of the brane tiling corresponding to regions in the Type IIB brane configuration occupied by stacks of $N$ D5-branes giving $U(N)_i$ gauge groups while the edges in the brane tiling correspond to chiral multiplets $X_{ij}$ that transform in the bifundamental representation of $(U(N)_i , \overline{U(N)}_j)$ corresponding to the adjacent faces of the edge.
Here, $i,j =1, \dots, G$ label the $U(N)$ gauge groups. 

The edges end at white and black nodes of the brane tiling, which have clockwise and anti-clockwise orientations, respectively.
This induces a natural ordering of edges ending at a given node of the brane tiling, which is interpreted 
as a gauge-invariant ordering of bifundamental chiral fields in the superpotential term corresponding to the brane tiling node.
For white nodes, the associated term is positive, while for black nodes the term is negative, as illustrated in \fref{f_fig00}. 
We call the dual graph of the brane tiling on $T^2$, replacing brane tiling faces to quiver nodes, edges to directed quiver edges, and white and black nodes to closed loops of quiver edges, as the \textit{periodic quiver} on $T^2$ \cite{Franco:2005rj}.

When we have $N=1$ such that all gauge groups are $U(1)$, the mesonic moduli space $\mathcal{M}^{mes}$ \cite{Feng:2000mi, Benvenuti:2006qr, Feng:2007ur, Butti:2007jv, Hanany:2012hi} of the $4d$ $\mathcal{N}=1$ supersymmetric gauge theory is given by the following symplectic quotient, 
\beal{es01a10}
\mathcal{M}^{mes} = \text{Spec} \left(\mathbb{C}[X_{ij}] / \langle \partial_{X_{ij}} W = 0 \rangle\right) // U(1)^{G-1}
~,~
\nn\\
\eea
where $\mathbb{C}[X_{ij}]$ is the coordinate ring formed by the set of bifundamental chiral fields $X_{ij}$, 
$ \mathcal{I} = \langle \partial_{X_{ij}} W = 0 \rangle$ is the ideal formed by the set of F-terms coming from the superpotential $W$, 
and $U(1)^{G-1}$ are the independent $U(1)$ gauge symmetries. 
In the case when $N=1$, $\mathcal{M}^{mes}$ is precisely the non-compact toric Calabi-Yau 3-fold probed by the D3-brane, whereas for $N>1$, the mesonic moduli space becomes the $N$-th symmetric product of the toric Calabi-Yau 3-fold. 

\begin{figure*}[ht!!]
\begin{center}
\resizebox{0.75\hsize}{!}{
\includegraphics[height=5cm]{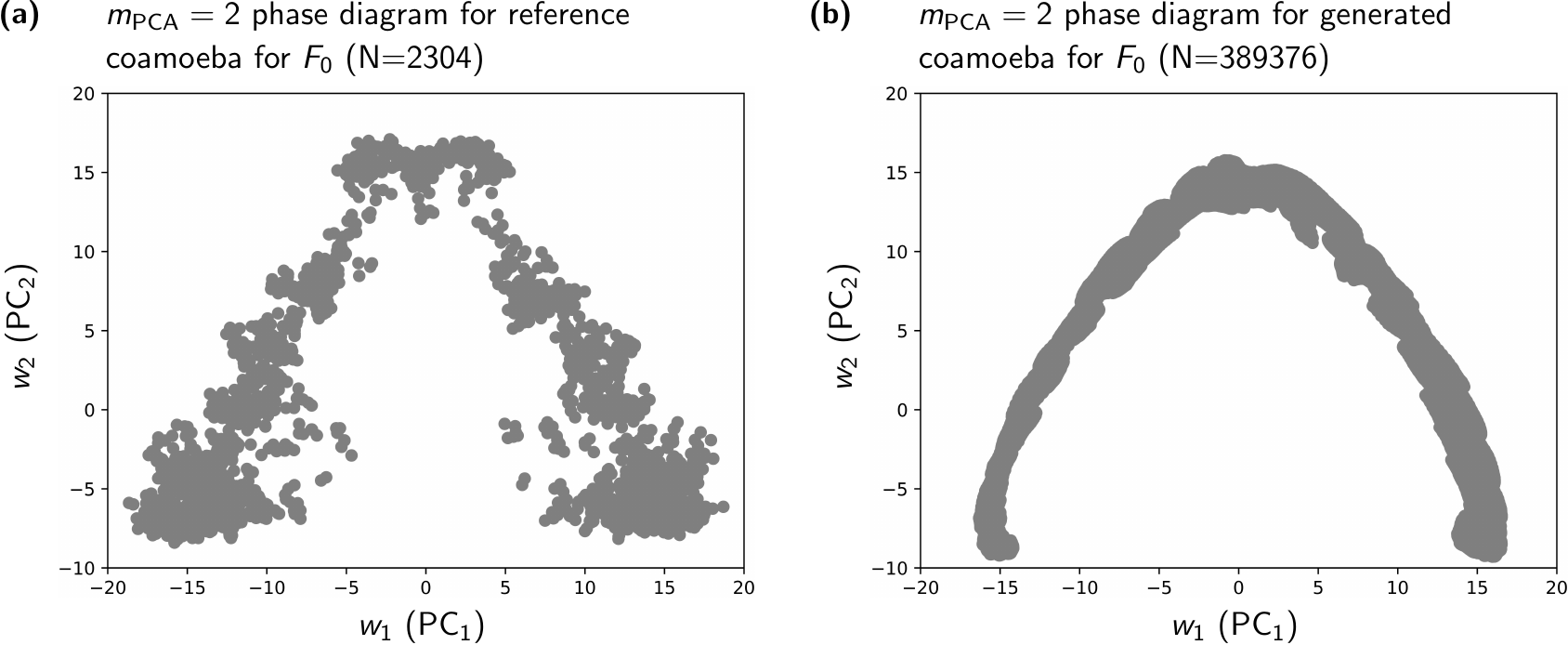} 
}
\caption{
(a) The $m_{\text{PCA}}=2$ dimensional phase diagram for $F_0$ obtained in \cite{Seong:2023njx} from $N=2304$ different coamoeba with complex structure moduli components $c_{ab} \in \{0, \pm 3, \pm 6, \pm 9\}$. 
Each point in the phase diagram represents a coamoeba with a unique choice of complex structure moduli. 
Using our trained CVA model, 
we are now able to generate many more coamoeba for $F_0$ within the same range of complex structure moduli. 
In (b), we obtain the $m_{\text{PCA}}=2$ dimensional phase diagram for $F_0$ based on $N=389376$ different coamoeba generated by our CVAE model with complex structure moduli components $c_{ab} \in \{0, \pm 0.5, \pm 1.0, \dots, \pm 5.5 ,\pm 6.0\}$.
The resulting phase diagram is such that points representing different coamoeba are much more closely placed by the PCA, giving a near-continuous phase diagram for $F_0$. 
\label{f_fig05}}
 \end{center}
 \end{figure*}

\section{Coamoeba Representation, Seiberg Duality and Phase Spaces \label{sec_coamoeba}}

The shape of the holomorphic curve $\Sigma$ depends on the values of the complex structure moduli $c_{(n_x, n_y)} \in \mathbb{C}^*$ in \eref{es01a00}. 
For different values of the complex structure moduli $c_{(n_x, n_y)}$, the shape of $\Sigma$ can change to such an extent that the associated Type IIB brane configurations undergo transitions that can be identified with \textit{Seiberg duality} \cite{Seiberg:1994pq, Feng:2000mi, Feng:2001bn, Feng:2001xr, Feng:2002zw} in terms of the realized $4d$ $\mathcal{N}=1$ supersymmetric gauge theories.
Such transitions under changes of complex structure moduli have also been studied extensively using Calabi-Yau mirror symmetry in \cite{Hori:2000ck, Hori:2000kt, horimirror, Franco:2016qxh, Franco:2016tcm}.

The $4d$ $\mathcal{N}=1$ supersymmetric gauge theories and brane tilings related to each other by Seiberg duality are referred to as \textit{toric phases} \cite{Feng:2000mi, Feng:2001bn, Feng:2001xr, Feng:2002zw}.
The transition from one toric phase to another toric phase in terms of brane tilings is given by a local deformation of the bipartite graph on $T^2$, which is also known as an \textit{urban renewal} or a \textit{spider move} \cite{goncharov2011dimers, ciucu1998complementation, kenyon1999trees}.
\fref{f_fig00} illustrates how this local deformation acts on the brane tilings.

It is challenging to identify which choice of 
complex structure moduli $c_{(n_x, n_y)}$ gives rise to which specific toric phase associated with a given toric Calabi-Yau 3-fold. 
Even though the values of $c_{(n_x, n_y)}$ determine the shape of the coamoeba on $T^2$ and the associated Type IIB brane configuration and brane tiling, it is challenging to identify at which values of $c_{(n_x, n_y)}$ transitions corresponding to Seiberg duality occur between toric phases.
This is especially the case when the number of complex structure moduli $c_{(n_x, n_y)}$ is large for certain toric Calabi-Yau 3-folds. 

Motivated by these questions, 
the work in \cite{Seong:2023njx}
employed \textit{unsupervised machine learning techniques}
in order to identify the toric phases for a toric Calabi-Yau 3-fold and moreover to generate a \textit{phase space} representation indicating the regions parameterized in terms of complex structure moduli that correspond to specific toric phases of the toric Calabi-Yau 3-fold.
By using techniques such as principal component analysis (PCA) \cite{pearson1901liii, hotelling1933analysis, jackson2005user, jolliffe2002principal, jolliffe1990principal, deisenroth2020mathematics} 
and $t$-distributed stochastic neighbor embedding ($t$-SNE) \cite{hinton2002stochastic, van2008visualizing},
the work was able to
to identify such a phase space
by dimensionally reducing the space of complex structure moduli $c_{(n_x, n_y)}$
and corresponding coamoeba.
\\

\begin{figure*}[ht!!]
\begin{center}
\resizebox{0.75\hsize}{!}{
\includegraphics[height=5cm]{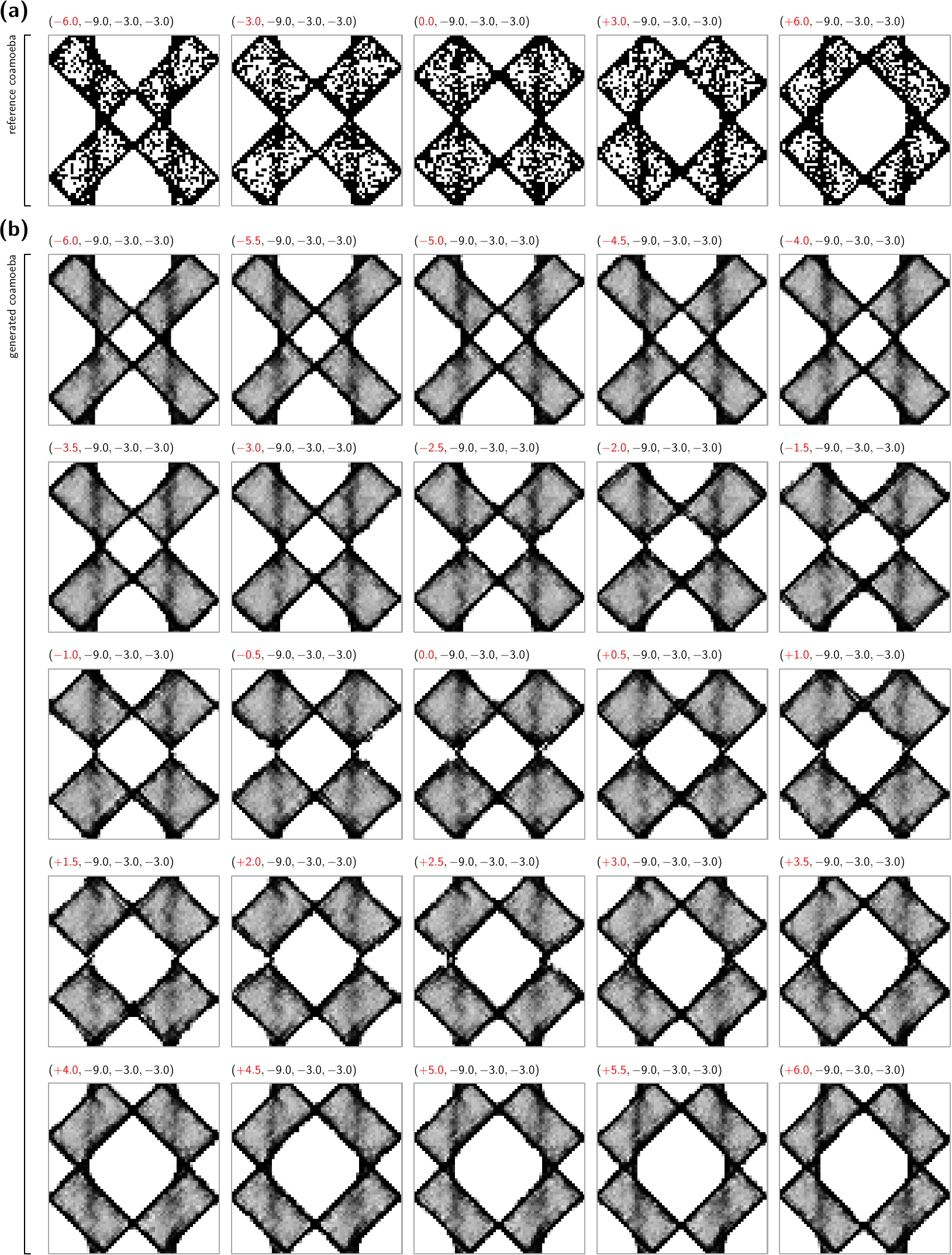} 
}
\caption{
(a) We have a selection of 5 coamoeba for $F_0$ with all having components of complex structure moduli $(c_{12}, c_{21}, c_{22})=(-9.0, -3.0, -3.0)$, whereas $-6.0 \leq c_{11} \leq +6.0$ is varied at coarse steps of $\Delta c_{11} = 3.0$. This sample of the training dataset is used to train the CVAE model, which after training is capable to generate 
(b) a selection of 25 coamoeba with $(c_{12}, c_{21}, c_{22})=(-9.0, -3.0, -3.0)$ and $-6.0 \leq c_{11} \leq +6.0$, where now $c_{11}$ can be varied at much smaller intervals $\Delta c_{11} = 0.5$. This includes coamoeba that were not part of the original training dataset and allows us to track infinitesimal changes in the coamoeba and its associated Type IIB brane configurations. 
\label{f_fig01}}
 \end{center}
 \end{figure*}

\noindent
\textit{Example.}
The work in \cite{Seong:2023njx} focused on the cone over
the zeroth Hirzebruch surface $F_0$, which is a toric Calabi-Yau 3-fold. 
The Newton polynomial for $F_0$ takes the form,
\beal{es01a20}
P(x,y) = \left(
x + \frac{1}{x}
\right)
+ c_1
\left(
y + \frac{1}{y}
\right)
+ c_2 ~,~
\eea
where the complex structure moduli are of the form $c_a = c_{a1} + i c_{a2} \in \mathbb{C}^*$.
By taking values $c_{ab} \in \{0, \pm 3, \pm 6, \pm 9\}$ such that $c_a \in \mathbb{C}^*$,
$N=2304$ different coamoeba were generated in \cite{Seong:2023njx}.
These were then via PCA dimensionally reduced to give a $m_{\text{PCA}}=2$ dimensional phase diagram as shown in \fref{f_fig05}(a).
Under $t$-SNE, 3 clusters of coamoeba were identified in \cite{Seong:2023njx} corresponding to the toric phases of $F_0$,
\beal{es01a21}
\ba{rl}
\text{phase 1} & \text{if $c_{11} = 0$}
\\
\text{phase 2a} & \text{if $c_{11} > 0$}
\\
\text{phase 2b} & \text{if $c_{11} < 0$}
\\
\ea
~.~
\eea
It was discovered that the sign of the real component $c_{11}$
determines the toric phase for $F_0$.
The reader is referred to \cite{Seong:2023njx} for more details about the use of PCA and $t$-SNE to identify the toric phases and phase diagram for toric Calabi-Yau 3-folds. 
\\

In this work,
we make use of the same techniques introduced in \cite{Seong:2023njx}
in order to plot the angular projection of $\Sigma ~:~ P(x,y) = 0$ onto $T^2$.
For the coamoeba plot on $T^2$, 
one needs to find first solutions for $P(x,y) = 0$.
These are then mapped to $T^2$ giving rise to the coamoeba plot.
In \cite{Seong:2023njx}, processes called $\theta_x$- and $\theta_y$-scan were introduced in order to find and map a random set of solutions for $P(x,y) = 0$ to $T^2$ for the coamoeba plot.

Let us give here a brief review of the $\theta_x$-scan.
Here, we pick a random angle $\theta_{x_0} \in [0, 2\pi)$
and a random radius $m_\epsilon \sim \mathcal{N}(0, \epsilon^2)$, 
where $\epsilon > 0$ is a small positive real number. 
These two random numbers allow us to define a randomly chosen $x$-component for a candidate solution for $P(x,y)=0$, which takes the general form
\beal{es01a22}
x_0 = e^{i\theta_{x_0} + m_\epsilon} ~.~
\eea
Using $x=x_0$, we have
\beal{es01a23}
P(x=x_0, y) 
= \sum_{(n_x, n_y) \in \Delta}
c_{(n_x,n_y)}
x_0^{n_x} y^{n_y} = 0 ~,~
\eea
which we can now solve for $y$. 
The solutions $y=y^*$
then can be combined with the randomly obtained value $x=x_0$ 
to give solutions to $P(x,y)=0$ mapped onto $T^2$, which take the form
\beal{es01a24}
(\theta_x^*, \theta_y^*)
=
(\theta_{x_0}, \arg(y^*))~.~
\eea
We can similarly define the $\theta_y$-scan process, where first a random value $\theta_{y_0} \in [0,2\pi)$ is picked in order to find 
solutions of the form $(\arg(x^*), \theta_{y_0})$.
The $\theta_x$- and $\theta_y$-scan can be done iteratively to find in total $N_\theta$ solutions for the coamoeba plot on $T^2$.
The higher the number $N_\theta$ of found solutions, the more resolved the coamoeba plot becomes. 

As in \cite{Seong:2023njx}, 
we discretize the coamoeba plot on $T^2$
for a set of solutions
of the form $(\theta_x^*, \theta_y^*)$
to the mirror curve $\Sigma ~:~ P(x,y) = 0$.
This discretized representation of the coamoeba plot is given by
a \textit{coamoeba matrix} $\mathbf{x}$, whose components are defined as follows,
\beal{es01a25}
x_{ij}
=
\left\{
\ba{cll}
1 &
\exists~ (\theta_x^*,\theta_y^*) ~s.t.~& (j-1)d_x \leq \theta_x^* < j d_x
\\
\; &
\hspace{1.5cm} and~&(i-1)d_y \leq \theta_y^* < i d_y
\\
0 &
\text{otherwise}
\ea
\right.
~,~
\nn\\
\eea
where $(\theta_x^*, \theta_y^*)$ is the angular part of a solution to $P(x,y) = 0$. 
The indices $i = 1, \dots, m_y$ and $j = 1, \dots, m_x$ label the grid points used for the discretized plot for the coamoeba.
Here, $0 < d_x \leq 2\pi$ and $0< d_y \leq 2\pi$ are the grid separation parameters given by
\beal{es01a26}
d_x = \frac{2\pi}{ m_x} ~,~
d_y = \frac{2\pi}{ m_y} ~.~ 
\eea

We can flatten the coamoeba matrix $\mathbf{x}$ such that it gives a \textit{coamoeba vector} of the following form, 
\beal{es01a26}
(x_{11}, \dots, x_{1 m_y}, x_{21}, \dots, x_{m_x 1}, \dots, x_{m_x m_y})
~.~
\eea
By construction, here the components of the coamoeba vector are $x_{ij} \in \{0,1\}$, but for the purpose of the remaining work we loosen this restriction such that $x_{ij} \in \mathbb{R}$.
In the following work, we make use of the coamoeba vector as a representation of the 
coamoeba projection of the mirror curve $\Sigma$.
By doing so, we are going to refer to both the coamoeba matrix and coamoeba vector as $\mathbf{x}$ and make the distinction explicit when we talk about their components and the number of indices labelling them. 
\\

\noindent
\textit{Example.}
\fref{f_fig01}(a) illustrates coamoeba plots for the zeroth Hirzebruch surface $F_0$
that have been generated with
parameters $\epsilon = 5$, $N_\theta = 2000$, and $m_x = m_y = 64$. 
For the remainder of this work, we will use the same parameters to generate coamoeba plots and their corresponding coamoeba matrices and vectors.
\\

\section{Generative AI: Conditional Variational Autoencoder (CVAE) \label{sec_cvae}}

In this work, we use a \textit{conditional variational autoencoder (CVAE)} \cite{kingma2013auto, sohn2015learning, cinelli2021variational} in order to generate coamoeba that play a quintessential role in determining corresponding Type IIB brane configurations.
These Type IIB brane configurations realize $4d$ $\mathcal{N}=1$ supersymmetric gauge theories associated to toric Calabi-Yau 3-folds.
The generated coamoeba are parameterized by the complex structure moduli in the mirror curve $\Sigma$ of the associated toric Calabi-Yau 3-fold.
A CVAE is a generative machine learning model that is used to generate data, which is conditioned on certain labels or attributes.
These are in our case the complex structure moduli labelling the coamoeba. 
CVAEs are extensions to the more standard \textit{variational autoencoder (VAE)} \cite{kingma2013auto} used commonly in modern machine learning. 
In the following section, we give a brief overview of CVAEs in relation to coamoeba and Type IIB brane configurations corresponding to toric Calabi-Yau 3-folds. 

\subsection{Latent Spaces and the Marginal Likelihood \label{sec_cvae_a}}

We describe in this section a conditional variational autoencoder (CVAE) \cite{kingma2013auto, sohn2015learning, cinelli2021variational}
that is trained to generate coamoeba $\mathbf{x} \in \mathbb{R}^{m_x \times m_y}$.
These coamoeba are labelled by complex structure moduli vectors of the form $\mathbf{c} \in (\mathbb{C}^*)^{|\Delta|-3}$ as introduced in \eref{es01a01b}, where the components are $c_a = c_{a1} + i c_{a2} \in \mathbb{C}^{*}$. 

The main objective of the CVAE in our work
is to generate coamoeba $\mathbf{x}$
by modelling and maximizing the \textit{marginal likelihood} $p(\mathbf{x}|\mathbf{c})$ of the form,
\beal{es05a01} 
p(\mathbf{x} | \mathbf{c})
= 
\int_L p(\mathbf{x}| \mathbf{z} , \mathbf{c} ) p(\mathbf{z} | \mathbf{c}) d\mathbf{z}
~,~
\eea
which gives the probability of observing a specific coamoeba $\mathbf{x}$ associated with the complex structure moduli $\mathbf{c}$. 
Here, $\mathbf{z}\in \mathbb{R}^m$ is the \textit{latent vector} parameterizing a $m$-dimensional \textit{latent space} $L$ over which the integral in \eref{es05a01} is taken. 

The role of the latent space vector $\mathbf{z}$
 is to capture features of the coamoeba $\mathbf{x}$
that are not represented by the actual values of the complex structure moduli $\mathbf{c}$. 
We can think of this latent space as an additional parameterization of the space of all possible coamoeba $\mathbf{x}$. 
The CVAE model uses this additional parameterization, for example, 
to capture the fact that certain choices of complex structure moduli $\mathbf{c}$ may result in the same or similar coamoeba $\mathbf{x}$. 
Accordingly, a role of the latent space vector $\mathbf{z}$ is to capture such \textit{multimodality} in the space of all possible coamoeba for a given mirror curve $\Sigma$. 
We also note here that the latent space vector $\mathbf{z}\in \mathbb{R}^m$ is chosen to be lower dimensional than the original coamoeba vectors $\mathbf{x}\in \mathbb{R}^{m_x \times m_y}$.

As defined in \eref{es05a01}, the marginal likelihood
$p(\mathbf{x} | \mathbf{c})$ can be expressed in terms of 
conditional probability distributions, which we summarize as follows:
\begin{itemize}
\item $p(\mathbf{x}| \mathbf{z} , \mathbf{c} )$: This is the \textit{likelihood} of the coamoeba $\mathbf{x}$ depending on the latent vector $\mathbf{z}$ and the complex structure moduli $\mathbf{c}$. 
Maximizing this probability ensures that the CVAE generates accurate coamoeba $\mathbf{x}$ based on a choice of latent vector $\mathbf{z}$ and complex structure moduli $\mathbf{c}$.
Given that the entries $x_{ij}$ of the coamoeba $\mathbf{x}$ are originally in $\{0,1\}$, we choose to approximate the likelihood to be a Bernoulli distribution of the following form, 
\beal{es05a04}
p(\mathbf{x}| \mathbf{z} , \mathbf{c} )
&=&
\prod_{i=1}^{m_x} \prod_{j=1}^{m_y} p(x_{ij} | \mathbf{z} , \mathbf{c} )
\nn\\
&=& 
\prod_{i=1}^{m_x} \prod_{j=1}^{m_y}
q_{\theta_d}(x_{ij}=1|\mathbf{z}, \mathbf{c})^{x_{ij}} 
\nn\\
&&
\hspace{0.15cm}
~\cdot~ (1- q_{\theta_d}(x_{ij}=1|\mathbf{z}, \mathbf{c}))^{1-x_{ij}} ~.~
\nn\\
\eea
Here,
$q_{\theta_d}(x_{ij}=1|\mathbf{z}, \mathbf{c})$
is the approximate probability for $p(x_{ij} = 1 | \mathbf{z} , \mathbf{c} )$, which is learned by the CVAE with model parameters $\theta_d$.
In the following work, we refer to these parameters as the \textit{decoder parameters} $\theta_d$. 

\item $p(\mathbf{z} | \mathbf{c})$: This is the \textit{prior distribution} over the latent variable $\mathbf{z}$, which depends on the values of the complex structure moduli $\mathbf{c}$. 
The choice of the prior over $\mathbf{z}$ allows us to model meaningful variations of the coamoeba $\mathbf{x}$ through the latent space $L$ parameterized by $\mathbf{z} \in \mathbb{R}^{m}$. 
For simplicity, we approximate in the following work the prior distribution to be independent of the complex structure moduli $\mathbf{c}$ and to be modeled by a standard normal distribution as follows,
\beal{es05a05}
p(\mathbf{z} | \mathbf{c})
\equiv
p(\mathbf{z})
=
\prod_{k=1}^{m}
\mathcal{N}(z_k; 0 , 1 )
~,~
\eea
where $z_k$ are the $m$ components of the latent space vector $\mathbf{z}$.
Here, the standard normal distribution is given by, 
\beal{es05a06}
\mathcal{N}(z_k; 0 , 1 )= 1/\sqrt{2\pi} \exp[- z_k^2/2] 
~.~
\eea

\end{itemize}

Even though the approximations that we introduce above for the likelihood $p(\mathbf{x}| \mathbf{z} , \mathbf{c} )$
and the prior distribution $p(\mathbf{x}| \mathbf{z} , \mathbf{c} )$ simplify the expression for the marginal likelihood in $p(\mathbf{x} | \mathbf{c})$ in \eref{es05a01}, 
it is still intractable to evaluate the integral over the whole latent space $L$, due the dimensionality of the latent space which we choose to be $m$ and the lack of analytical methods to evaluate the integral over the latent space.
In the following sections, we circumvent this obstacle by introducing the posterior distribution, which allows us to rewrite the integral under \textit{variational inference} \cite{jordan1999introduction, attias1999variational, blei2017variational}.
\\

\subsection{Bayes' Theorem and Posterior Distribution \label{sec_cvae_b}}

Under \textit{Bayes' theorem} \cite{doi:10.1098/rstl.1763.0053}, we can relate the likelihood of the coamoeba $p(\mathbf{x}|\mathbf{z}, \mathbf{c})$ and the prior distribution $p(\mathbf{z}|\mathbf{c})$ over the latent variable $\mathbf{z} \in L$ to the \textit{posterior distribution} $p(\mathbf{z}|\mathbf{x}, \mathbf{c})$, which takes the following form,
\beal{es06a05}
p(\mathbf{z}|\mathbf{x}, \mathbf{c})
= 
\frac{
p(\mathbf{x}|\mathbf{z},\mathbf{c}) p(\mathbf{z}|\mathbf{c})
}{
p(\mathbf{x}|\mathbf{c})
} ~,~
\eea
where we also make use of the marginal likelihood $p(\mathbf{x}|\mathbf{c})$ from \eref{es05a01}. 
The role of the posterior distribution $p(\mathbf{z}|\mathbf{x}, \mathbf{c})$ is to represent the updated probability distribution of the latent variable $\mathbf{z}$ based on the observed coamoeba $\mathbf{x}$ and the corresponding complex structure moduli $\mathbf{c}$.
It ensures that the CVAE model learns during training a meaningful latent space representation of the essential features in the coamoeba data, including its multimodality under the complex structure moduli $\mathbf{c}$.

As a probability density function, the posterior distribution $p(\mathbf{z}|\mathbf{x}, \mathbf{c})$
is normalized and its integral over the entire latent space $L$ gives, 
\beal{es06a10}
\int_{L} p(\mathbf{z} | \mathbf{x}, \mathbf{c}) d\mathbf{z}
=
1 ~.~
\eea
This can be seen when we rewrite the posterior distribution in terms of the expression for the marginal likelihood in \eref{es05a01}, which gives
\beal{es06a11}
p(\mathbf{z}|\mathbf{x}, \mathbf{c})
= 
\frac{
p(\mathbf{x}|\mathbf{z},\mathbf{c}) p(\mathbf{z}|\mathbf{c})
}{
\int_L p(\mathbf{x}|\mathbf{z},\mathbf{c}) p(\mathbf{z}|\mathbf{c}) d\mathbf{z}
}~,~
\eea
where we can see that the right hand side becomes $1$ when the posterior distribution $p(\mathbf{z}|\mathbf{x}, \mathbf{c})$ is integrated over the entire latent space $L$.
Another way to interpret \eref{es06a11} is that the marginal likelihood $p(\mathbf{x}|\mathbf{c})$ and its expression in \eref{es05a01} plays the role of normalizing the posterior distribution $p(\mathbf{z}|\mathbf{x}, \mathbf{c})$.

\begin{figure*}[ht!!]
\begin{center}
\resizebox{0.95\hsize}{!}{
\includegraphics[height=5cm]{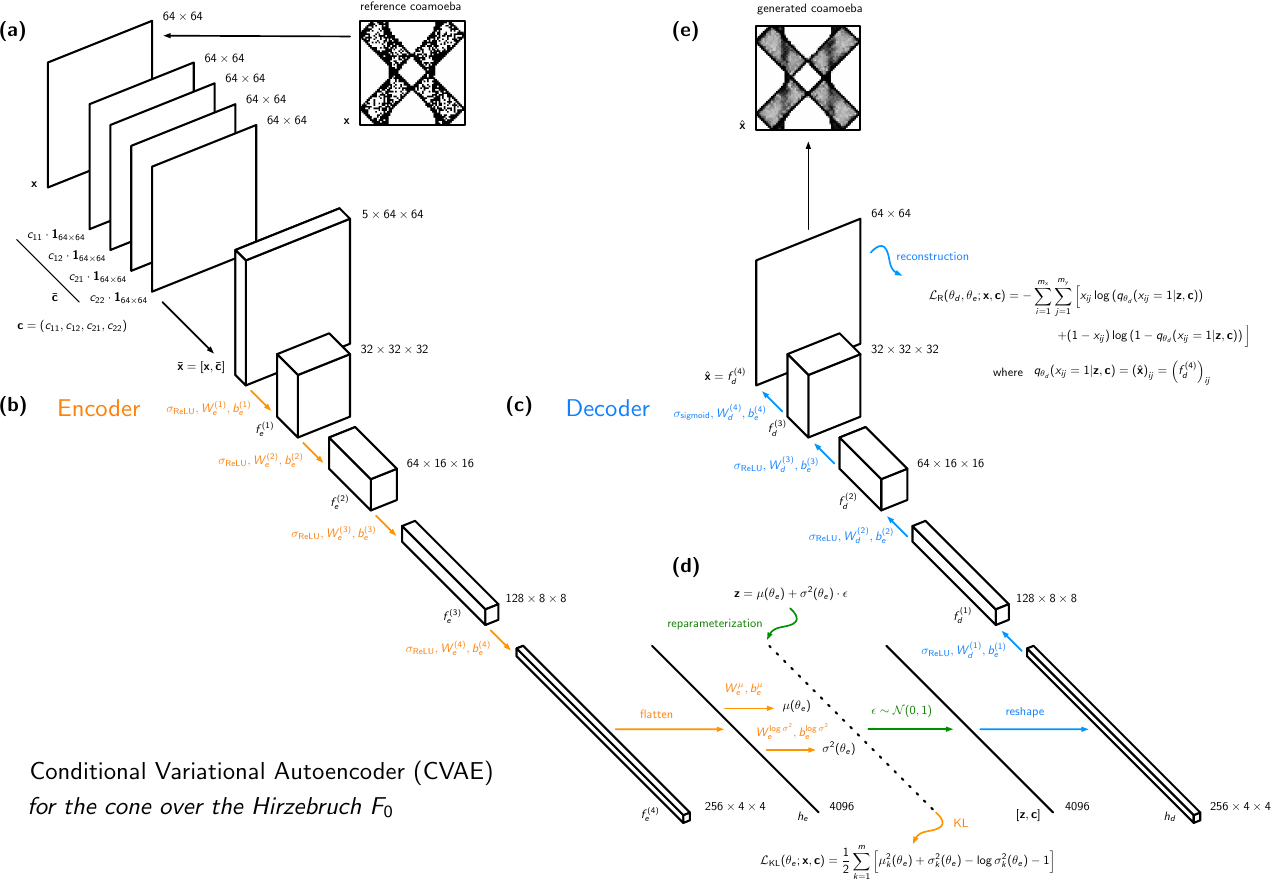} 
}
\caption{
(a) The coamoeba given by $\mathbf{x}$ and the associated complex structure moduli $\mathbf{c}$
are combined into a tensor $\overline{\mathbf{x}}$ of dimension $5\times 64\times 64$, which is the input tensor for the conditional variational autoencoder (CVAE) model illustrated here. 
(b) The encoder network of the CVAE model consists of a fully-connected convolutional neural network with 4 layers $f_e^{(u)}(\theta_e)$, where the output layer is then connected to a fully-connected linear layer that outputs the mean $\mu_k(\theta_e)$ and logarithmic variance $\log \sigma_k^2(\theta_e)$ of the approximated posterior distribution $q_{\theta_e}(\mathbf{z}|\mathbf{x}, \mathbf{c})$, where $k=1, \dots, m$ and $m$ is the dimension of the latent space $L$. 
(c) The decoder network of the CVAE model also consists of a fully-connected convolutional neural network with 4 layers $f_d^{(v)}(\theta_d)$, which outputs the probability distribution for the likelihood in terms of $q_{\theta_d}(x_{ij}=1 | \mathbf{z}, \mathbf{c})$. This measures the probability of the coamoeba occupying grid point $(i,j)$ in the discretized unit cell of $T^2$. 
(d) The decoder takes as an input the reparameterized latent space vector $\mathbf{z}$
and (e) outputs what we refer to as the generated coamoeba $\hat{\mathbf{x}}$ corresponding to $\mathbf{z}$ and a choice of complex structure moduli $\mathbf{c}$.
\label{f_fig04}}
 \end{center}
 \end{figure*}

\subsection{Maximizing the Marginal Likelihood and the Evidence Lower Bound \label{sec_cvae_c}}

The CVAE has the aim to maximize the marginal likelihood $p(\mathbf{x}|\mathbf{c})$ defined in \eref{es05a01}.
This ensures that it becomes more likely to observe from a trained CVAE the correct coamoeba given by $\mathbf{x} \in \mathbb{R}^{m_x \times m_y}$ for a choice of complex structure moduli $\mathbf{c} \in (\mathbb{C}^*)^{|\Delta|-3}$.

As mentioned above, the computation of the marginal likelihood $p(\mathbf{x}|\mathbf{c})$ based on \eref{es05a01} is not possible, because it involves integrating over the whole $m$-dimensional latent space $L$.
This is where the posterior distribution $p(\mathbf{z}|\mathbf{x}, \mathbf{c})$ enables us to simplify the expression for the marginal likelihood without having to directly compute the integral over the latent space $L$.

In the context of our CVAE, it is equivalent to consider the \textit{marginal log-likelihood} $\log p(\mathbf{x}|\mathbf{c})$, which can be written as follows,
\beal{es06a15}
\log p(\mathbf{x} | \mathbf{c})
&=&
\log
\int_L
p(\mathbf{x}, \mathbf{z} | \mathbf{c})
d\mathbf{z}
\nn\\
&=&
\log \int_L
q_{\theta_e}(\mathbf{z}| \mathbf{x}, \mathbf{c}) 
\frac{
p(\mathbf{x}, \mathbf{z}| \mathbf{c})
}{
q_{\theta_e}(\mathbf{z}| \mathbf{x}, \mathbf{c}) 
}
d\mathbf{z}
~.~
\nn\\
\eea
Above, we introduce an \textit{approximate posterior distribution} $q_{\theta_e}(\mathbf{z}|\mathbf{x}, \mathbf{c})$ in order to model the true posterior distribution $p(\mathbf{z}|\mathbf{x}, \mathbf{c})$.
This is because the definition of the posterior distribution $p(\mathbf{z}| \mathbf{x}, \mathbf{c})$ from \eref{es06a05} involves the same integral over the latent space $L$ that we want to avoid as in the case for the marginal likelihood $p(\mathbf{x}|\mathbf{c})$ in \eref{es05a01}. 
We recall that the likelihood $p(\mathbf{x}| \mathbf{z}, \mathbf{c})$ in \eref{es05a04} is approximated for our CVAE model with a Bernoulli distribution, which depends on an approximate probability $q_{\theta_d}(x_{ij}=1|\mathbf{z}, \mathbf{c})$
and the decoder parameters $\theta_d$ of our CVAE model. 
For the posterior distribution $p(\mathbf{z}|\mathbf{x}, \mathbf{c})$, we make use of a similar approximation in terms of $q_{\theta_e}(\mathbf{z}|\mathbf{x}, \mathbf{c})$, which depends on the \textit{encoder parameters} $\theta_e$ of our CVAE model.
The ultimate aim of the CVAE model therefore can here be summarized as finding the optimal parameters $\theta_d^*$ and $\theta_e^*$ such that the marginal log-likelihood $\log p(\mathbf{x} | \mathbf{c})$ is maximized in \eref{es06a15}. 

Let us quantify the optimal model parameters $\theta_d^*$ and $\theta_e^*$ in our CVAE model.
We first note that the expected value of a given function $f(\mathbf{z})$ under a probability distribution $p(\mathbf{z})$ can be written as follows, 
\beal{es06a20}
E_{p(\mathbf{z})} [f(\mathbf{z})] = 
\int_L 
p(\mathbf{z}) f(\mathbf{z}) d\mathbf{z}
~,~
\eea
where above we have $\mathbf{z}$ as our latent space vector and $L$ as our latent space. 
Based on this, we identify in the expression for the marginal log-likelihood in \eref{es06a15}
that we have an expression for an expected value under the probability distribution given by the approximate posterior distribution $q_{\theta_e}(\mathbf{z}|\mathbf{x}, \mathbf{c})$, 
\beal{es06a22}
\log p(\mathbf{x}|\mathbf{c})
= 
\log E_{q_{\theta_e}(\mathbf{z} | \mathbf{x}, \mathbf{c})} \left[
\frac{
p(\mathbf{x}, \mathbf{z}| \mathbf{c})
}{
q_{\theta_e}(\mathbf{z}| \mathbf{x}, \mathbf{c}) 
}
\right]
~.~
\eea

Under
\textit{Jensen's inequality} \cite{jensen1906fonctions} given by,
\beal{es06a25}
f(E[\mathbf{z}]) \geq E[f(\mathbf{z})]
~,~
\eea
we can rewrite the expression for the marginal log-likelihood in \eref{es06a22} in terms of a lower bound as follows,
\beal{es06a26}
\log p(\mathbf{x}|\mathbf{c}) 
&\geq&
\int q_{\theta_e}(\mathbf{z} | \mathbf{x}, \mathbf{c})
\log 
\left(
\frac{
p(\mathbf{x}, \mathbf{z}| \mathbf{c})
}{
q_{\theta_e}(\mathbf{z}| \mathbf{x}, \mathbf{c}) 
}
\right)
\nn\\
&=&
E_{q_{\theta_e}(\mathbf{z} | \mathbf{x}, \mathbf{c})} \left[
\log
\left(
\frac{
p(\mathbf{x}, \mathbf{z}| \mathbf{c})
}{
q_{\theta}(\mathbf{z}| \mathbf{x}, \mathbf{c}) 
}
\right)
\right]
~.~
\nn\\
\eea
Given that the joint conditional probability $p(\mathbf{x}, \mathbf{z} | \mathbf{c})$ can be written as a product under the chain rule,  
\beal{es06a26b}
p(\mathbf{x}, \mathbf{z} | \mathbf{c}) = p(\mathbf{x} | \mathbf{z} , \mathbf{c}) \cdot p(\mathbf{z}| \mathbf{c})
~,~
\eea
the lower bound for the marginal log-likelihood in \eref{es06a26} can be expressed as follows, 
\beal{es06a30}
\log p(\mathbf{x}|\mathbf{c})
&\geq&
E_{q_{\theta_e}(\mathbf{z} | \mathbf{x}, \mathbf{c})}[
\log p(\mathbf{x}|\mathbf{z}, \mathbf{c})
]
\nn\\
&&
+
E_{q_{\theta_e}(\mathbf{z} | \mathbf{x}, \mathbf{c})}[
\log p(\mathbf{z}|\mathbf{c})
]
\nn\\
&&
-
E_{q_{\theta_e}(\mathbf{z} | \mathbf{x}, \mathbf{c})}[
\log q_{\theta_e}(\mathbf{z}|\mathbf{x},\mathbf{c})
]
~.~
\nn\\
\eea
The above lower bound on the marginal log-likelihood $\log p(\mathbf{x}|\mathbf{c})$ is known as the \textit{evidence lower bound (ELBO)} \cite{kingma2013auto}, which we can see depends only on expected values under $q_{\theta_e}(\mathbf{z}|\mathbf{x},\mathbf{c})$ for the likelihood $p(\mathbf{x}| \mathbf{z} , \mathbf{c} )$ depending on $q_{\theta_d}(x_{ij}=1|\mathbf{z}, \mathbf{c})$ in \eref{es05a04}, the prior distribution $p(\mathbf{z}|\mathbf{c})$ chosen to be approximated by the standard normal distribution in \eref{es05a05}, and finally the approximate posterior distribution $q_{\theta_e}(\mathbf{z}|\mathbf{x}, \mathbf{c})$.

We can see here that the primary aim of the CVAE model is to maximize the ELBO
by optimizing $q_{\theta_e}(\mathbf{z}|\mathbf{x}, \mathbf{c})$ and $q_{\theta_d}(x_{ij}=1|\mathbf{z}, \mathbf{c})$.
In the following section, 
we summarize this optimization process in the CVAE model.

\begin{figure*}[ht!!]
\begin{center}
\resizebox{0.85\hsize}{!}{
\includegraphics[height=5cm]{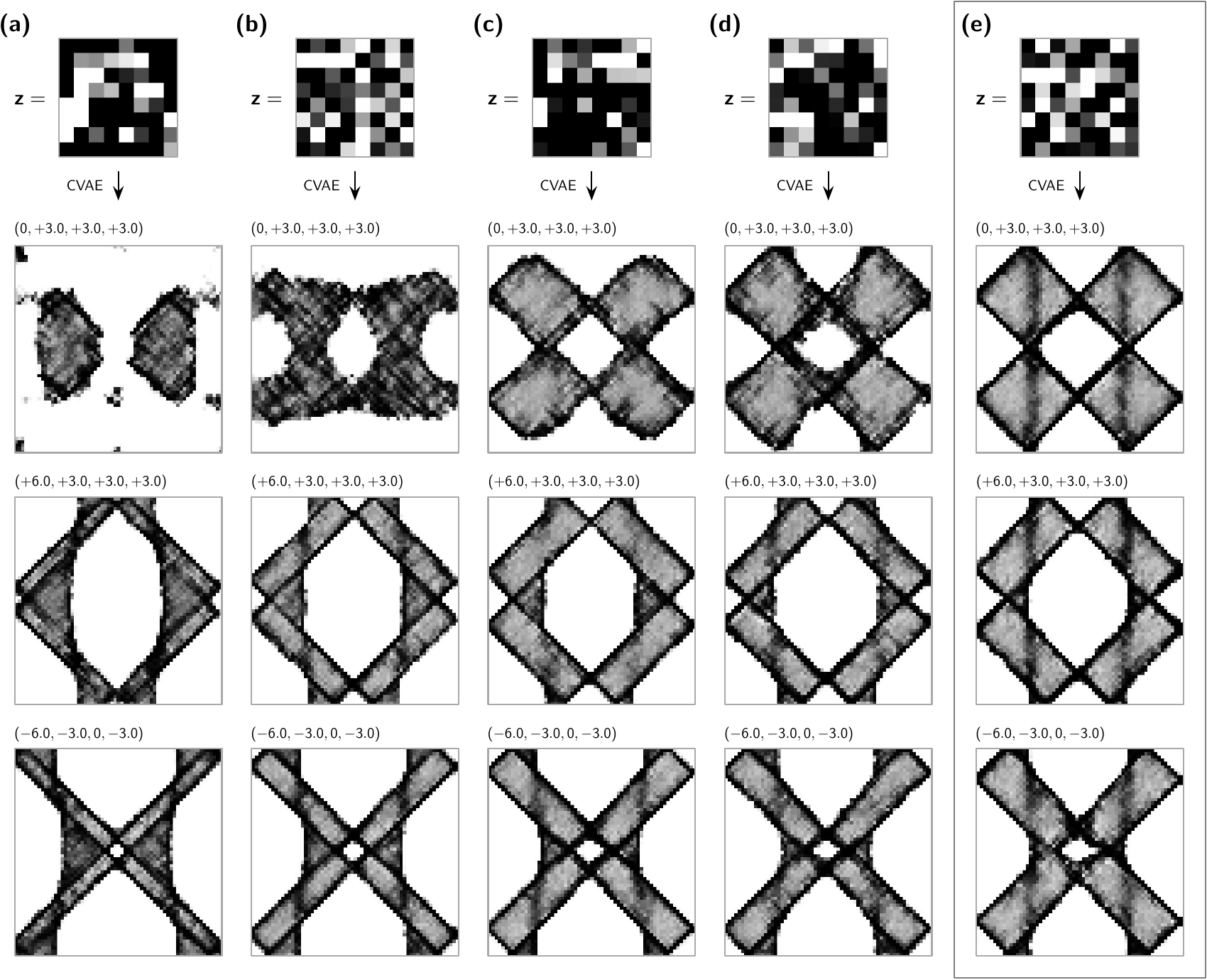} 
}
\caption{
(a)-(d) Certain choices of latent space vectors $\mathbf{z}$
are less optimal than other choices for coamoeba generated by the CVAE model. 
The choice in (e) is what we use throughout this work as the optimal choice for the latent space vector $\mathbf{z}$.
The optimal choice for $\mathbf{z}$ appears to produce coamoeba on $T^2$ with sharp boundaries.
Given that the CVAE model outputs a probability map on $T^2$ given in terms of $q_{\theta_d}(x_{ij}=1|\mathbf{z}, \mathbf{c})$, 
the less optimal choices for $\mathbf{z}$ appear to generate probability distributions that are more spread out, 
reflecting an increased level of uncertainty in the locality of the coamoeba on $T^2$ and the branes in the corresponding Type IIB brane configuration. 
\label{f_fig02}}
 \end{center}
 \end{figure*}

\subsection{Encoder, Decoder and Reparameterization \label{sec_cvae_d}}

The negative ELBO serves as the \textit{loss function} $\mathcal{L}$ for the CVAE model, and takes the following form
\beal{es07a10}
&&
\mathcal{L}(\theta_e, \theta_d; \mathbf{x}, \mathbf{c})
=
-
E_{q_{\theta_e}(\mathbf{z} | \mathbf{x}, \mathbf{c})}[
\log p(\mathbf{x}|\mathbf{z}, \mathbf{c})
]
\nn\\
&&
\hspace{0.4cm}
-
\Big[
E_{q_{\theta_e}(\mathbf{z} | \mathbf{x}, \mathbf{c})}[
\log p(\mathbf{z}|\mathbf{c})
]
\nn\\
&&
\hspace{0.8cm}
-
E_{q_{\theta_e}(\mathbf{z} | \mathbf{x}, \mathbf{c})}[
\log q_{\theta_e}(\mathbf{z}|\mathbf{x},\mathbf{c})
]
\Big]
~.~
\eea
As in all machine learning models, the loss function here quantifies the difference between the expected output and the actual output of the machine learning model. 
Accordingly, our CVAE model has the aim to minimize the loss function $\mathcal{L}$ during the training process. 

The CVAE model that we are using in this work is made of 3 main components that each
play a role in minimizing the loss function $\mathcal{L}$ in \eref{es07a10}.
The following section summarizes the role and architecture of each of these components.
\\

\noindent
\textbf{Encoder.}
The first component of the CVAE model is the 
\textit{encoder}, which has the main aim to minimize the last two terms of the loss function $\mathcal{L}$ in \eref{es07a10}.
This part of the loss function is also known as the \textit{Kullback-Leibler (KL) divergence term} \cite{10.1214/aoms/1177729694, kullback1997information}, given by
\beal{es07a15}
&&
\mathcal{L}_{\text{KL}}(\theta_e; \mathbf{x}, \mathbf{c})
=
-
\Big[
E_{q_{\theta_e}(\mathbf{z} | \mathbf{x}, \mathbf{c})}[
\log p(\mathbf{z}|\mathbf{c})
]
\nn\\
&&
\hspace{2.8cm}
-
E_{q_{\theta_e}(\mathbf{z} | \mathbf{x}, \mathbf{c})}[
\log q_{\theta_e}(\mathbf{z}|\mathbf{x},\mathbf{c})
]
\Big]
~.~
\nn\\
\eea
In statistics, the Kullback-Leibler divergence is a measure of how 
a reference probability distribution, in our case the prior distribution $p(\mathbf{z}|\mathbf{c})$,
is different from a second probability distribution, which in our case is the approximate posterior distribution $q_{\theta_e}(\mathbf{z}|\mathbf{x}, \mathbf{c})$.
By minimizing the KL divergence term, 
we are making sure that the approximate posterior distribution approaches the prior distribution.

Recalling that in section \sref{sec_cvae_a}, we approximated the prior distribution $p(\mathbf{z}|\mathbf{c})$ with a standard normal distribution, we approximate here also the posterior distribution $q_{\theta_e}(\mathbf{z}|\mathbf{x}, \mathbf{c})$ with a standard normal distribution, where now its mean $\mu_k(\theta_e)$ and variance $\sigma_k^2(\theta_e)$ has to be optimized by the encoder.
Using the expressions of the prior distribution and the approximate posterior distribution,
\beal{es07a16}
p(\mathbf{z}|\mathbf{c})
&=& \prod_{k=1}^{m} \mathcal{N}(z_k; 0,1)
~,~
\nn\\
q_{\theta_e}(\mathbf{z}|\mathbf{x},\mathbf{c})
&=& \prod_{k=1}^{m} \mathcal{N}(z_k; \mu_k(\theta_e), \sigma_k^2(\theta_e))
~,~
\eea
we can write the KL divergence term as follows, 
\beal{es07a18}
&&
\mathcal{L}_{\text{KL}}(\theta_e; \mathbf{x}, \mathbf{c})
=
\frac{1}{2}
\sum_{k=1}^{m}
\Big[
\mu_k^2(\theta_e)
+
\sigma_k^2(\theta_e)
\nn\\
&&
\hspace{3cm}
- 
\log \sigma_k^2(\theta_e)
- 1
\Big]
~,~
\eea
which we can see is in terms of the mean $\mu_k(\theta_e)$ and variance $\sigma_k^2(\theta_e)$
of the standard normal distribution for the approximate posterior distribution $q_{\theta_e}(\mathbf{z}|\mathbf{x}, \mathbf{c})$.

The task for the encoder is to obtain the optimal mean $\mu_k(\theta_e)$ and variance $\sigma_k^2(\theta_e)$ for $q_{\theta_e}(\mathbf{z}|\mathbf{x}, \mathbf{c})$ that minimizes the KL divergence term of the loss function $\mathcal{L}$ for our CVAE model.
This is done by training the encoder on a set of coamoeba $\mathbf{x}$ and the corresponding complex structure moduli $\mathbf{c}$ for a given toric Calabi-Yau 3-fold. 
In order to ensure a simple model architecture with a single input channel, we first define an input tensor $\bar{\mathbf{x}}$ that combines the coamoeba $\mathbf{x}$ with its complex structure moduli $\mathbf{c}$ as follows,
\beal{es07a20}
\bar{\mathbf{x}}= [\mathbf{x}, \bar{\mathbf{c}} ] \in \mathbb{R}^{(2|\Delta|-5) \times m_x \times m_y}
\eea
where each of the complex structure moduli given by the components of $\mathbf{c}$
are lifted to $m_x \times m_y$ matrices, 
\beal{es07a19}
\bar{\mathbf{c}}= \mathbf{c} \cdot \mathbf{1}_{m_x \times m_y} \in \mathbb{R}^{2(|\Delta|-3) \times m_x \times m_y}
~.~
\eea
Here, $\mathbf{1}_{m_x \times m_y}$ is a $m_x \times m_y$ matrix with all of its entries set to 1. 
We further note that for the encoder, the coamoeba $\mathbf{x}$ with its label $\mathbf{c}$
is combined into a tensor with dimension $(2|\Delta|-5) \times m_x \times m_y$, where
$|\Delta|$ corresponds to the number of vertices in the toric diagram of the toric Calabi-Yau 3-fold. 
\\

\noindent
\textit{Example.}
In our example of the zeroth Hirzebruch surface $F_0$ with $|\Delta|=5$, 
whose mirror curve is given in \eref{es01a20},
the input tensor $\bar{\mathbf{x}}$ would be of dimension $5 \times m_x \times m_y$.
The independent components of the complex structure moduli vector $\mathbf{c}$ would be $(c_{11}, c_{12}, c_{21}, c_{22}) \in \mathbb{R}^4$. 
As indicated in section \sref{sec_coamoeba}, 
we take the dimension of the coamoeba matrices $\mathbf{x}$ to be $m_x = m_y = 64$, and assume this to be the case 
for the remainder of this work without loss of generality.
\\

Taking the input tensor $\bar{\mathbf{x}} \in \mathbb{R}^{(2|\Delta|-5) \times m_x \times m_y}$, 
the encoder is a \textit{fully-connected convolutional neural network} \cite{lecun1989backpropagation, lecun1995learning, lecun1998gradient}, 
defined as follows, 
\beal{es07a21}
f_e^{(1)}(\theta_e)  &=&
\sigma_{\text{ReLU}} \left(
W_e^{(1)}\cdot \bar{\mathbf{x}} + b_e^{(1)}
\right)
\in \mathbb{R}^{32 \times 32 \times 32}
\nn\\
f_e^{(2)}(\theta_e) &=&
\sigma_{\text{ReLU}} \left(
W_e^{(2)} \cdot f_e^{(1)} + b_e^{(2)}
\right)
\in \mathbb{R}^{64 \times 16 \times 16}
\nn\\
f_e^{(3)}(\theta_e) &=&
\sigma_{\text{ReLU}} \left(
W_e^{(3)}\cdot f_e^{(2)} + b_e^{(3)}
\right)
\in \mathbb{R}^{128 \times 8 \times 8}
\nn\\
f_e^{(4)}(\theta_e) &=&
\sigma_{\text{ReLU}} \left(
W_e^{(4)} \cdot f_e^{(3)} + b_e^{(4)}
\right)
\in \mathbb{R}^{256 \times 4 \times 4}
~,~
\nn\\
\eea
where
$f_e^{(u)}(\theta_e)$ is the output tensor for the layers of the encoder network labelled by $u=1, \dots, 4$.
Each layer has tensorial weights $W_e^{(u)}$ and biases $b_e^{(u)}$ that are all part of the encoder parameters $\theta_e$.
The activation function used for each layer is the \textit{rectified linear unit (ReLU)} $\sigma_{\text{ReLU}}$ \cite{householder1941theory, fukushima1969visual, nair2010rectified}, which takes the following tensorial form, 
\beal{es07a21b}
\sigma_{\text{ReLU}}(\mathbf{x})
= \max{(\mathbf{0}, \mathbf{x})}
~.~
\eea

The last part of the encoder takes the output of the fully-connected convolutional neural network, 
which is a tensor with dimension $256 \times 4 \times 4$, and flattens it to a vector of the form
\beal{es07a22}
h_e = \text{flatten}(f_e^{(4)}) \in \mathbb{R}^{4096} ~.~
\eea
It is then taken into 
a fully-connected linear layer that outputs the mean $\mu_k(\theta_e)$
and logarithmic variance $\log \sigma^2_k(\theta_e)$
of the approximated posterior distribution $q_{\theta_e}(\mathbf{z}|\mathbf{x}, \mathbf{c})$ as follows,
\beal{es07a23}
\mu(\theta_e) 
&=&
\Big(
W^{\mu}_e \cdot h_e + b^\mu_e
\Big) \in \mathbb{R}^{m}
~,~
\nn\\
\log\sigma^2(\theta_e) 
&=&
\Big(
W^{\log\sigma^2}_e \cdot h_e + b^{\log\sigma^2}_e
\Big) \in \mathbb{R}^{m}
~,~
\nn\\
\eea
where the tensorial weights $W^{\mu}_e$ and $W^{\log\sigma^2}_e$, and biases $b^\mu_e$ and $b^{\log\sigma^2}_e$
are also part of the encoder parameters $\theta_e$.
Here, we note that $m$ is the dimension of the latent space $L$
parameterized by latent space vector $\mathbf{z} \in \mathbb{R}^m$.

The encoder has the purpose of minimizing the KL divergence term in the loss function $\mathcal{L}$ for the CVAE model
by optimizing the encoder parameters $\theta_e$.
This in turn leads to the optimal values for the mean and variance for the approximated posterior distribution $q_{\theta_e}(\mathbf{z}|\mathbf{x}, \mathbf{c})$, making it close to the prior distribution $p(\mathbf{z}|\mathbf{c})$ in the CVAE.
Moreover, minimizing the KL divergence term
achieves a \textit{regularization of the latent space $L$} \cite{kingma2013auto, sohn2015learning}, 
by preventing overfitting of the training data 
and by structuring the latent space meaningfully such that similar coamoeba are mapped to close latent vectors in the latent space $L$.
\\

\noindent
\textit{Example.}
In our example of the zeroth Hirzebruch surface $F_0$, 
we take the dimension of the latent space $L$ to be $m=64$ without loss of generality.
We note here that the dimension of the latent space is a hyperparameter 
of the CVAE model.
\\

\noindent
\textbf{Reparameterization.}
The \textit{reparameterization} \cite{kingma2013auto, sohn2015learning} step in the CVAE model acts as a differentiable bridge between the encoder and what we later introduce as the decoder of the CVAE model. 
It expresses the latent space vector $\mathbf{z} \in \mathbb{R}^m$
in terms of the mean $\mu_k(\theta_e)$
and standard deviation $\sigma_k(\theta_e)$ of the approximate posterior distribution $q_{\theta_e}(\mathbf{z}|\mathbf{x}, \mathbf{c})$ from the encoder as follows,
\beal{es07a24}
z_k = \mu_k(\theta_e) + \sigma_k(\theta_e) \cdot \varepsilon
~,~
\eea
where $k=1, \dots, m$
and $\varepsilon$
is a noise variable sampled from $\mathcal{N}(0,1)$
and independent of the encoder parameters $\theta_e$.

We can see that the latent space vector $\mathbf{z} \in \mathbb{R}^m$
becomes differentiable in terms of the encoder parameters $\theta_e$ due to the reparameterization in \eref{es07a24}. 
This allows us to define a gradient of the latent vector $\mathbf{z}$
with respect to the encoder parameters $\theta_e$.
Such a gradient is essential for training the encoder network and the CVAE model as a whole via \textit{backpropagation} \cite{ackley1985learning, rumelhart1986learning, lecun1988theoretical} -- an algorithm
used to train neural networks through the computation of gradients of the loss function $\mathcal{L}$ with respect to neural network parameters. 
Backpropagation would be not possible if we would directly sample stochastically $\mathbf{z} \sim q_{\theta_e}(\mathbf{z}|\mathbf{x}, \mathbf{c})$, making the reparameterization in \eref{es07a24} essential for training the CVAE model. 
\\

\noindent
\textbf{Decoder.}
The final component of the CVAE model is the \textit{decoder}, which has the main aim to minimize the first term of the CVAE loss function $\mathcal{L}$ in \eref{es07a10}. 
This part of the loss function is known as the
\textit{reconstruction loss term} and is given by, 
\beal{es07a25}
\mathcal{L}_{\text{R}}(\theta_d, \theta_e; \mathbf{x}, \mathbf{c})
=
-
E_{q_{\theta_e}(\mathbf{z} | \mathbf{x}, \mathbf{c})}[
\log p(\mathbf{x}|\mathbf{z}, \mathbf{c})
]~,~
\eea
where $q_{\theta_e}(\mathbf{z}|\mathbf{x}, \mathbf{c})$
is the approximate posterior distribution from the encoder, and 
$p(\mathbf{x}|\mathbf{z}, \mathbf{c})$ is the likelihood for the coamoeba $\mathbf{x}$. 
The reconstruction loss term ensures that the CVAE model
generates outputs $\hat{\mathbf{x}}$ that accurately represent the original coamoeba $\mathbf{x}$ used to train the CVAE model for a given choice of complex structure moduli $\mathbf{c}$.
The objective of the decoder is to minimize the reconstruction loss, which corresponds to maximizing the likelihood $p(\mathbf{x}|\mathbf{z}, \mathbf{c})$ of obtaining coamoeba $\mathbf{x}$ for a given choice of complex structure moduli $\mathbf{c}$.

We recall that in \eref{es05a04}, 
we approximated the likelihood with a Bernoulli distribution depending on
$q_{\theta_d}(x_{ij} = 1|\mathbf{z}, \mathbf{c})$, which is the approximate probability for the coamoeba component $x_{ij} = 1$.
The decoder learns $q_{\theta_d}(x_{ij} = 1|\mathbf{z}, \mathbf{c})$ while minimizing 
the reconstruction loss term, which now in terms of the Bernoulli distribution in \eref{es05a04} can be written as, 
\beal{es07a30}
&&
\mathcal{L}_{\text{R}}(\theta_d, \theta_e; \mathbf{x}, \mathbf{c})
=
- \sum_{i=1}^{m_x} \sum_{j=1}^{m_y} 
\Big[
x_{ij} \log\left( q_{\theta_d}( x_{ij} = 1 | \mathbf{z}, \mathbf{c}) \right)
\nn\\
&&
\hspace{1cm}
+
(1-x_{ij}) \log\left(
1- 
q_{\theta_d}( x_{ij} = 1 | \mathbf{z}, \mathbf{c})
\right)
\Big]
~.~
\eea
Here, we note that the reconstruction loss term depends on both the encoder parameters $\theta_e$ and the decoder parameters $\theta_d$.
While the probability function $q_{\theta_d}( x_{ij} = 1 | \mathbf{z}, \mathbf{c})$
for the Bernoulli distribution depends on the
decoder parameters $\theta_d$
and has to be obtained through the decoder, 
the dependency on the encoder parameters $\theta_e$
originates from the latent space vector $\mathbf{z}$
which is taken as an input for the decoder using the reparameterization in \eref{es07a24}. 

\begin{figure*}[ht!!]
\begin{center}
\resizebox{0.85\hsize}{!}{
\includegraphics[height=5cm]{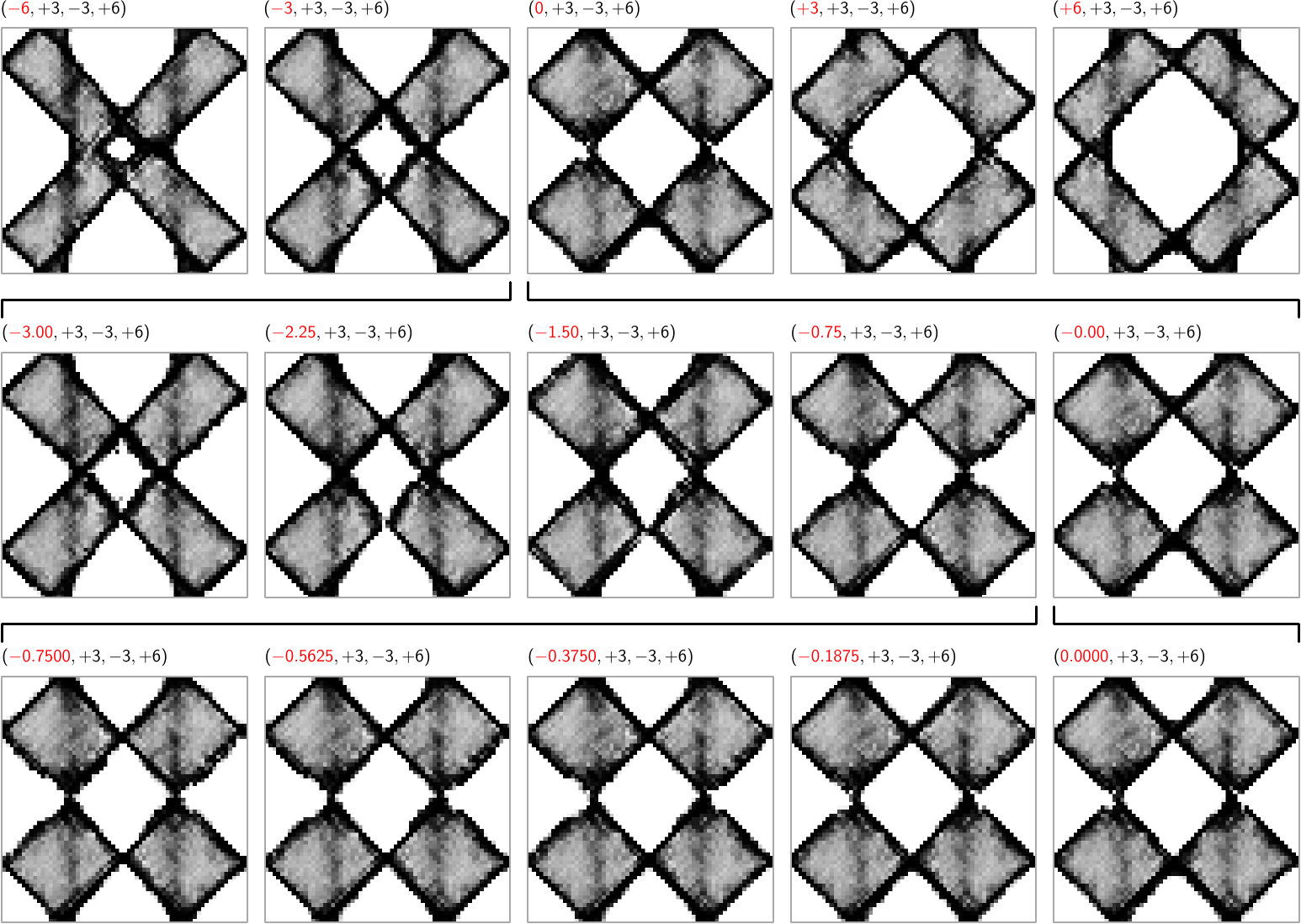} 
}
\caption{
The trained CVAE model is able to generate coamoeba 
that are closer from each other in terms of their corresponding complex structure moduli. 
This allows us to generate with the trained CVAE model a near-continuous 
representation of the transitions between coamoeba and the corresponding Type IIB brane configurations. 
These near-continuous sequences of transitions between coamoeba are represented as phase paths
in the phase diagram of a given toric Calabi-Yau 3-fold. 
\label{f_fig06}}
 \end{center}
 \end{figure*}

Combined with the complex structure moduli $\mathbf{c}$, 
we can define the input for the decoder as the following concatenated vector, 
\beal{es07a31}
[\mathbf{z}, \mathbf{c}] \in \mathbb{R}^{m + 2(|\Delta|-3) }
~.~
\eea
This $m + 2(|\Delta|-3)$ - dimensional vector can be reshaped into a 3-tensor of the following form, 
\beal{es07a32}
h_d = \text{reshape}([\mathbf{z}, \mathbf{c}]) \in \mathbb{R}^{m_a \times m_b \times m_c}
\eea
using what is known as a 
row-major order mapping 
given by, 
\beal{es07a32}
(h_d)_{abc} = [\mathbf{z}, \mathbf{c}]_{(a\times m_b \times m_c) + (b\times m_c) + c}
~,~
\eea
where $a=1, \dots , m_a$, $b=1, \dots , m_b$, and $c=1, \dots, m_c$.
Given that the output $f^{(4)}_e(\theta_e) \in \mathbb{R}^{256\times 4\times 4}$ of the encoder network in \eref{es07a21}
is a 3-tensor of dimension $256\times 4\times 4$, 
we choose here
$m_a = 256$, $m_b=4$, $m_c =4$.

Taking the input tensor $h_d \in \mathbb{R}^{256\times 4\times 4}$, 
the decoder is a fully-connected convolutional neural network 
with the aim of identifying the optimal probability function $q_{\theta_d}( x_{ij} = 1 | \mathbf{z}, \mathbf{c})$
for the Bernoulli distribution 
while minimizing the reconstruction loss term $\mathcal{L}_{\text{R}}$ in \eref{es07a25}.
The decoder network takes the following form, 
\beal{es07a33}
f_d^{(1)}(\theta_d)
&=& 
\sigma_{\text{ReLU}} \left(
W^{(1)}_d \cdot h_d + b^{(1)}_d 
\right) \in \mathbb{R}^{128 \times 8 \times 8}
\nn\\
f_d^{(2)}(\theta_d)
&=& 
\sigma_{\text{ReLU}} \left(
W^{(2)}_d \cdot f_d^{(1)} + b^{(2)}_d
\right) \in \mathbb{R}^{64 \times 16 \times 16}
\nn\\
f_d^{(3)}(\theta_d)
&=& 
\sigma_{\text{ReLU}} \left(
W^{(3)}_d \cdot f_d^{(2)} + b^{(3)}_d
\right) \in \mathbb{R}^{32 \times 32 \times 32}
\nn\\
f_d^{(4)}(\theta_d)
&=& 
\sigma_{\text{sigmoid}} \left(
W^{(4)}_d \cdot f_d^{(3)} + b^{(4)}_d
\right) \in \mathbb{R}^{1 \times 64\times 64}
~,~
\nn\\
\eea
where $f^{(v)}_d (\theta_d)$
is the output tensor for the decoder layers labelled by $v=1, \dots, 4$.
As in the encoder network, each decoder layer has tensorial weights $W_d^{(v)}$ and biases $b_d^{(v)}$
that are part of the decoder parameters $\theta_d$.
The activation function for the first three layers is $\sigma_{\text{ReLU}}$,
whereas the last layer has a \textit{sigmoid activation function} $\sigma_{\text{sigmoid}}$ \cite{han1995influence}, which is defined as follows, 
\beal{es07a33b}
\sigma_{\text{sigmoid}}(\mathbf{x})
= 
\frac{1}{1+e^{-\mathbf{x}}}
~.~
\eea
Here, we note that the sigmoid activation function has a range in $[0,1]$, 
which is chosen such that its output can be interpreted as a probability.

In fact, the output $f_d^{(4)}$ of the decoder network is as expected 
interpreted as the probability $q_{\theta_d}( x_{ij} = 1 | \mathbf{z}, \mathbf{c})$.
Given that the output is a 2-tensor of dimension $64\times 64$, the same dimensions for the discretized coamoeba on $T^2$
with $m_x = m_y = 64$, 
we can interpret the probability map $p_{ij} = q_{\theta_d}( x_{ij} = 1 | \mathbf{z}, \mathbf{c})$
as a representation of the generated coamoeba given 
latent space vector $\mathbf{z}$ and choice of complex structure moduli $\mathbf{c}$.
In summary, we have therefore
\beal{es07a34}
q_{\theta_d}( x_{ij} = 1 | \mathbf{z}, \mathbf{c})
=
\left(f_d^{(4)}\right)_{ij}
=
\left(\hat{\mathbf{x}}\right)_{ij}
~,~
\eea
where $i,j = 1, \dots, 64$
and $\hat{\mathbf{x}}$ is the generated coamoeba by the CVAE model for given latent vector $\mathbf{z}$ and choice of complex structure moduli $\mathbf{c}$.

\subsection{Gradients and CVAE Training}

In order to minimize the overall loss function given by
\beal{es07a40}
\mathcal{L}(\theta_e, \theta_d; \mathbf{x}, \mathbf{c})
=
\mathcal{L}_{\text{KL}}(\theta_e; \mathbf{x}, \mathbf{c})
+
\mathcal{L}_{\text{R}}(\theta_d, \theta_e; \mathbf{x}, \mathbf{c})
~,~
\nn\\
\eea
with respect to the encoder parameters $\theta_e$ and decoder parameters $\theta_d$, 
we use \textit{gradient-based optimization} \cite{ackley1985learning, rumelhart1986learning, lecun1988theoretical}.
The model parameters take the following form, 
\beal{es07a35}
\theta_e 
&=&
\left\{
W_e^{(u)}, b_e^{(u)},
W_e^{\mu}, b_e^{\mu},
W_e^{\log\sigma^2}, b_e^{\log\sigma^2}
\right\} ~,~
\nn\\
\theta_d 
&=&
\left\{
W_d^{(v)}, 
b_d^{(v)}
\right\}
~,~
\eea
where $u,v=1, \dots, 4$ label the encoder and decoder network layers, respectively. 

As part of this optimization process, 
we have \textit{backpropagation} \cite{ackley1985learning, rumelhart1986learning, lecun1988theoretical},
which effectively 
computes gradients of the loss function with respect to $\theta_e$ and $\theta_d$
and backward passes the gradients through each network layer.
During this process, 
backpropagation updates the encoder and decoder parameters in order to minimize
the reconstruction loss and KL divergence, effectively training the CVAE model through gradient descent.
The parameter updates are done as follows,
\beal{es07a36}
\theta_e^\prime = \theta_e - 
\eta
\frac{\partial \mathcal{L}}{\partial \theta_e}
~,~
\theta_d^\prime = \theta_d - 
\eta
\frac{\partial \mathcal{L}}{\partial \theta_d}
~,~
\eea
where $\eta$ is the learning rate. 

In this section, we summarize the gradients
that play a key role in the optimization process in the CVAE model for coamoeba. 
We first have a look at the gradients in terms of the encoder parameters $\theta_e$, 
which take the following form,
\beal{es07a40}
&&
\frac{
\partial \mathcal{L}(\theta_e, \theta_d; \mathbf{x}, \mathbf{c})
}{
\partial
\theta_e
}
=
\frac{\partial
\mathcal{L}_{\text{KL}}(\theta_e; \mathbf{x}, \mathbf{c})
}{
\partial\theta_e}
+
\frac{\partial
\mathcal{L}_{\text{R}}(\theta_d, \theta_e; \mathbf{x}, \mathbf{c})
}{
\partial\theta_e}
~,~
\nn\\
\eea
where the $\theta_e$-dependency is in both the KL divergence term $\mathcal{L}_{\text{KL}}$ and the reconstruction loss term $\mathcal{L}_{\text{R}}$ of the loss function $\mathcal{L}$ in \eref{es07a40}.
Recalling the expression for the KL divergence term $\mathcal{L}_{\text{KL}}$ in \eref{es07a18}, 
we can write the corresponding gradient in terms of $\theta_e$ as follows,
\beal{es07a41}
&&
\frac{\partial
\mathcal{L}_{\text{KL}}(\theta_e; \mathbf{x}, \mathbf{c})
}{
\partial\theta_e
}
=
\frac{1}{2}
\sum_{k=1}^{m} \Bigg[
\frac{\partial
\mathcal{L}_{\text{KL}}(\theta_e; \mathbf{x}, \mathbf{c})
}{
\partial\mu_k(\theta_e)
}
\cdot
\frac{
\partial\mu_k(\theta_e)
}{
\partial\theta_e
}
\nn\\
&&
\hspace{2cm}
+
\frac{\partial
\mathcal{L}_{\text{KL}}(\theta_e; \mathbf{x}, \mathbf{c})
}{
\partial\sigma_k(\theta_e)
}
\cdot
\frac{
\partial\sigma_k(\theta_e)
}{
\partial\theta_e
}
\Bigg]
~,~
\eea
where
\beal{es07a42}
\frac{\partial
\mathcal{L}_{\text{KL}}(\theta_e; \mathbf{x}, \mathbf{c})
}{
\partial\mu_k(\theta_e)
}
&=&
\mu_k(\theta_e) ~,~
\nn\\
\frac{\partial
\mathcal{L}_{\text{KL}}(\theta_e; \mathbf{x}, \mathbf{c})
}{
\partial\sigma_k(\theta_e)
}
&=&
\sigma_k(\theta_e) - \frac{1}{\sigma_k(\theta_e)}
~.~
\eea
The gradient for the reconstruction loss term $\mathcal{L}_{\text{R}}$ in \eref{es07a30}
in terms of $\theta_e$ takes the following form,
\beal{es07a43}
&&
\frac{\partial
\mathcal{L}_{\text{R}}(\theta_d, \theta_e; \mathbf{x}, \mathbf{c})
}{
\partial\theta_e}
=
\sum_{i=1}^{m_x} \sum_{j=1}^{m_y}
\frac{
\partial\mathcal{L}_{\text{R}}(\theta_d, \theta_e; \mathbf{x}, \mathbf{c})
}{
\partial \hat{{x}}_{ij}
}
\nn\\
&&
\hspace{4cm}
\cdot
\frac{
\partial \hat{{x}}_{ij}
}{
\partial \mathbf{z}
}
\cdot
\frac{
\partial \mathbf{z}
}{
\partial \theta_e
}
~,~
\eea
where
\beal{es07a44}
\frac{
\partial\mathcal{L}_{\text{R}}(\theta_d, \theta_e; \mathbf{x}, \mathbf{c})
}{
\partial \hat{{x}}_{ij}
}
&=&
- \frac{x_{ij}}{\hat{{x}}_{ij}}
+ \frac{1- x_{ij}}{1-\hat{{x}}_{ij}}
~,~
\nn\\
\frac{
\partial z_k
}{
\partial \theta_e
}
&=&
\frac{
\partial \mu_k(\theta_e)
}{
\partial \theta_e
}
+ 
\frac{
\partial \sigma_k(\theta_e)
}{
\partial \theta_e
}
\cdot
\varepsilon
~.~
\nn\\
\eea
Here, we note that the 
derivates
$\frac{
\partial \mu_k(\theta_e)
}{
\partial \theta_e
}
$
and
$\frac{
\partial \sigma_k(\theta_e)
}{
\partial \theta_e
}$
are
obtained via backpropagation along the encoder network with layers $f^{(u)}(\theta_e)$ in \eref{es07a21}
and the fully connected linear layers for the mean $\mu(\theta_e)$ and log-variance $\log \sigma^2(\theta_e)$ in \eref{es07a23}.
Similarly, the derivative $\frac{
\partial \hat{{x}}_{ij}
}{
\partial \mathbf{z}
}$
is obtained via backpropagation along the decoder and encoder networks, where $\hat{{x}}_{ij}$
is given in \eref{es07a34}.

The gradient in terms of the decoder parameters $\theta_d$ can be found as follows, 
\beal{es07a50}
&&
\frac{
\partial \mathcal{L}(\theta_e, \theta_d; \mathbf{x}, \mathbf{c})
}{
\partial
\theta_d
}
=
\frac{\partial
\mathcal{L}_{\text{R}}(\theta_d, \theta_e; \mathbf{x}, \mathbf{c})
}{
\partial\theta_d}
~,~
\eea
where we note that only the reconstruction loss term $\mathcal{L}_{\text{R}}$ has dependency on the decoder parameters $\theta_d$.
Recalling the reconstruction loss term in \eref{es07a30}, we can write
\beal{es07a51}
&&
\frac{\partial
\mathcal{L}_{\text{R}}(\theta_d, \theta_e; \mathbf{x}, \mathbf{c})
}{
\partial\theta_d}
=
\sum_{i=1}^{m_x} \sum_{j=1}^{m_y}
\frac{
\partial\mathcal{L}_{\text{R}}(\theta_d, \theta_e; \mathbf{x}, \mathbf{c})
}{
\partial \hat{{x}}_{ij}
}
\cdot
\frac{
\partial \hat{{x}}_{ij}
}{
\partial \theta_d
}
~,~
\nn\\
\eea
where we have as before, 
\beal{es07a52}
\frac{
\partial\mathcal{L}_{\text{R}}(\theta_d, \theta_e; \mathbf{x}, \mathbf{c})
}{
\partial \hat{{x}}_{ij}
}
&=&
- \frac{x_{ij}}{\hat{{x}}_{ij}}
+ \frac{1- x_{ij}}{1-\hat{{x}}_{ij}}
~,~
\eea
and the derivative $\frac{
\partial \hat{{x}}_{ij}
}{
\partial \theta_d
}$ is obtained through backpropagation along the decoder network
with layers $f^{(v)}(\theta_d)$ in \eref{es07a33}.
\\

\begin{figure*}[ht!!]
\begin{center}
\resizebox{0.98\hsize}{!}{
\includegraphics[height=5cm]{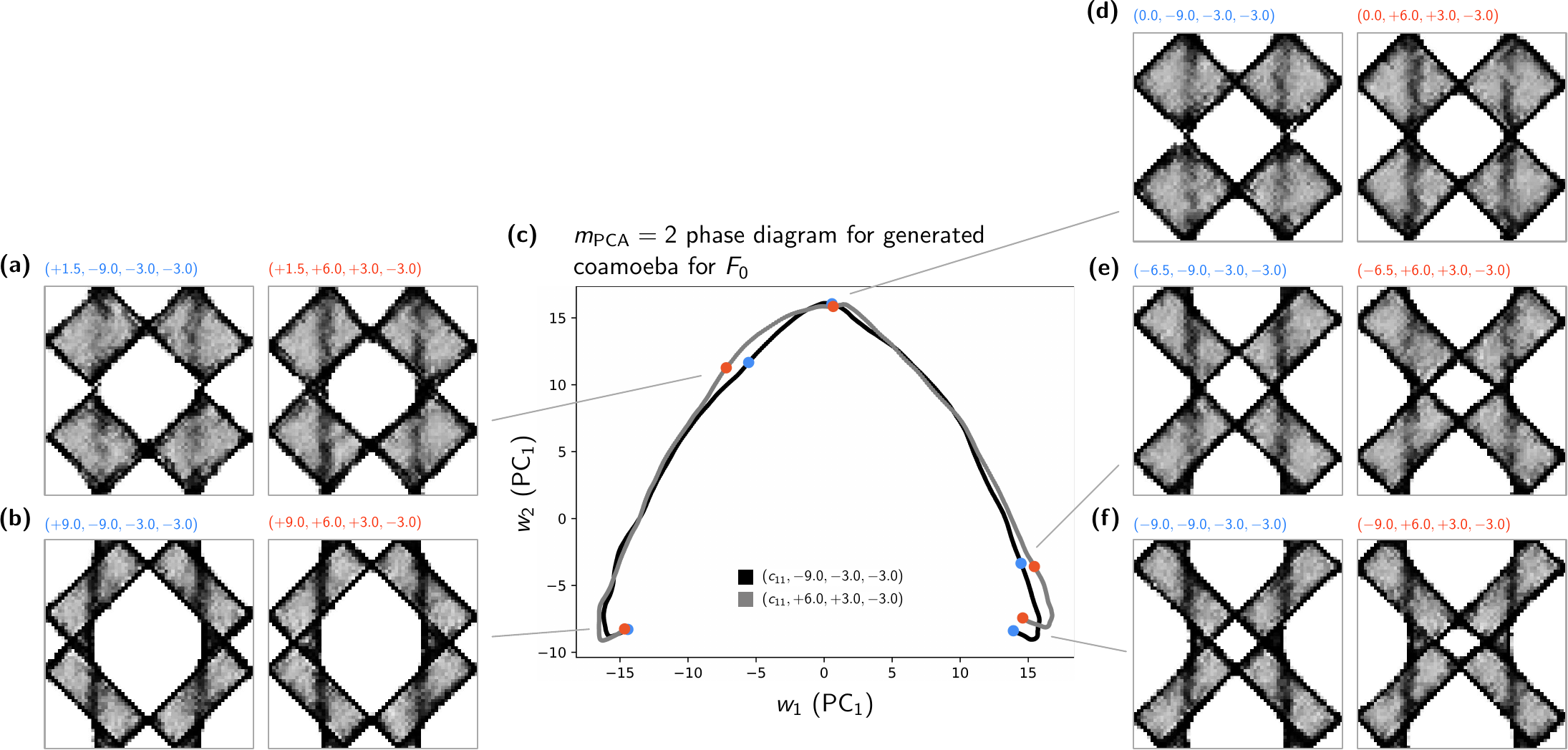} 
}
\caption{
Two sample phase paths $\rho_{(+6,+3,-3)})(t)$ [gray] and $\rho_{(-9,-3,-3)})(t)$ [black]
represented in the $m_{\text{PCA}}=2$ dimensional phase space of $F_0$ in (c). 
The order parameter $t=c_{11}$ for $F_0$ is taken for both paths to be in $-9.0 \leq t \leq +9.0$
and each path consists of $721$ coamoeba generated at intervals of $\Delta t =0.05$.
Snapshots of coamoeba for both phase paths at (a) $t=+1.5$, (b) $t=+9.0$, (d) $t=0.0$, (e) $t=-6.5$ and (f) $t=-9.0$ are shown. 
\label{f_fig03}}
 \end{center}
 \end{figure*}

\section{New Brane Configurations \\ from Old and Discussions}

The CVAE model that we introduce in this work is expected to 
generate the coamoeba for any toric Calabi-Yau 3-fold
and corresponding choice of complex structure moduli. 
In this work, we illustrate the capabilities of this generative AI model for 
the coamoeba and the corresponding Type IIB brane configurations for the cone over the zeroth Hirzebruch surface $F_0$ \cite{Beasley:1999uz, Feng:2000mi}.
We can report that the CVAE model has been tested successfully on other toric Calabi-Yau 3-folds, which is going to be the subject of a future publication.
We can comment here that for other toric Calabi-Yau 3-folds, the only adaptation of the CVAE model architecture arises due to the different number of complex structure moduli $\mathbf{c}$ for different toric Calabi-Yau 3-folds in \eref{es07a20}.
\\

\noindent
\textit{Example.}
Using the mirror curve in \eref{es01a20} for $F_0$
and a selection of complex structure moduli $c_{a1}+i c_{a2} \in \mathbb{C}^*$
with $c_{ab}\in \{0,\pm 3, \pm 6, \pm 9\}$, 
we obtain a training set of $N=2304$ different coamoeba and their corresponding coamoeba matrices $\mathbf{x}$
with $\epsilon = 5$, $N_\theta = 2000$ and $m_x=m_y = 64$.
As introduced in section \sref{sec_cvae} and illustrated in \fref{f_fig04}, 
we employ a CVAE model and train it on this dataset with $1000$ epochs that each have batches of size $32$. 
Here, we recall that the batch size is the number of samples in the training dataset which the model processes before it updates its model parameters.
The epoch number refers to the total number of times the model goes through the entire training dataset during training. 

We can confirm that after training, the CVAE model is capable of generating the coamoeba for any choice of complex structure moduli $\mathbf{c}$ for the zeroth Hirzebruch surface $F_0$.
Given that the output $\hat{\mathbf{x}}$ of the CVAE model is the probability distribution of the following form, 
\beal{es08a01}
q_{\theta_d}(x_{ij}=1|\mathbf{z}, \mathbf{c}) = (\hat{\mathbf{x}})_{ij}
~,~
\eea
the generated coamoeba depends not just on the choice of complex structure moduli $\mathbf{c}$
but also the latent vector $\mathbf{z} \in \mathbb{R}^{64}$.
\fref{f_fig01}(b)
illustrates a collection of 25 coamoeba generated for a range of complex structure moduli $\mathbf{c}$
and a fixed latent space vector of the form, 
\beal{es08a02}
&&
\mathbf{z}=
\nn\\
&&
\left[
{\tiny
\ba{cccccccc}
-0.26 , &  +1.54 , &  +0.05 , &  +0.65 , & -0.75 , & -0.25 , &  +1.07 , &  +0.27 , \\  
+ 0.09 , & +0.05 , & +0.68 , & -0.94 , & -0.70  , &  +0.63 , & -0.29 , & -1.36 , \\  
+1.56 , &  +1.78 , & +1.21 , &  +0.29 , & +1.95 , &  +0.85 , &  +0.52 , & -0.46 , \\ 
-1.19 , &  +0.86 , & -0.98 , & +0.87 , &  +1.10  , & -1.87 , & -0.64 , &  +0.56 , \\  
+0.08 , &  +1.87 , &  +0.27 , &  +0.67 , & -0.39 , & -0.59 , &  +2.62 , &  +0.06 , \\  
+1.05 , & -0.48 , &  +0.87 , & -0.57 , &  +0.28 , & +0.90  , & -0.56 , & -0.36 , \\ 
-1.51 , & -0.20  , &  +1.12 , &  +0.52 , & -0.33 , & -0.33 , & +0.97 , &  +0.28 , \\
-0.60  , &  +0.59 , & -0.31 , & +0.03 , &  +0.14 , & -0.53 , &  +0.32 , & +0.19
\ea
}
\right
]
~.~
\nn\\
\eea
When testing the trained generative AI model, we will use the canonical choice in \eref{es08a02}
for the latent space vector $\mathbf{z}$ unless otherwise indicated. 
We comment here that the study of the structure and properties of the latent space of a trained CVAE model is very interesting and is going to be the subject of future work.
Furthermore, the choice of the optimal latent vector $\mathbf{z}$ will fundamentally differ between different toric Calabi-Yau 3-folds, which are going to be studied using CVAE models in future work.

\fref{f_fig02} illustrates generated coamoeba for different choices of latent space vectors $\mathbf{z}$.
The choice in \eref{es08a02} corresponds to \fref{f_fig02}(e)
giving a sharp boundary for the coamoeba on $T^2$.
We observe here that the CVAE model introduces with its latent space \textit{new moduli} in terms of $\mathbf{z}$.
These moduli appear
to not just parameterize the multimodality of the coamoeba mentioned in section \sref{sec_cvae_a}, but also
appear to measure the boundary sharpness of the mirror curve projected onto $T^2$
as well as the branes involved in the corresponding Type IIB brane configuration. 
\\

Let us summarize some of the capabilities of the trained CVAE model for $F_0$
by the following overview of new observations in this work:
\\

\paragraph{Continuous Transitions between coamoeba.}
The trained CVAE model allows us to generate an arbitrary number of new coamoeba within a given finite interval in the space of all complex structure moduli.
This property is illustrated in \fref{f_fig06}. 
As a consequence of this property, the trained CVAE model appears to have learned a complete representation of the mirror curve $\Sigma$ at every
value of the complex structure moduli.

Because of this, the CVAE model also knows about the Type IIB brane configurations
consisting of the D5-branes suspended between the NS5-brane wrapping $\Sigma$
at all values of the complex structure moduli for $\Sigma$.
We can use this property of the generative AI model to track the changes of $\Sigma$, deformations of the coamoeba projection on $T^2$
and movements of branes in the resulting Type IIB brane configurations
during continuous variations of the complex structure moduli.
As a result of this new capability, we are able to track using the CVAE model
continuous transitions between distinct toric phases of the corresponding $4d$ $\mathcal{N}=1$ supersymmetric gauge theories.
\\

\paragraph{Continuous Phase Spaces.}
As illustrated in \cite{Seong:2023njx}, 
a collection of coamoeba at different values of complex structure moduli
can be dimensionally reduced using unsupervised machine learning techniques such as PCA or $t$-SNE
to obtain a low-dimensional representation of the phase space of the associated $4d$ $\mathcal{N}=1$ supersymmetric gauge theories.
As observed in \cite{Seong:2023njx}, 
such phase space
representations contain clustered regions that correspond to different toric phases of the corresponding $4d$ $\mathcal{N}=1$ supersymmetric gauge theories. 

In \cite{Seong:2023njx}, a $m_{\text{PCA}}=2$ dimensional phase space was obtained with a collection of $N=2304$ coamoeba at different values of the complex structure moduli for the zeroth Hirzebruch surface $F_0$.
This $m_{\text{PCA}}=2$ phase diagram is reproduced in \fref{f_fig05}(a).
Given that our trained CVAE model is capable of generating any number of coamoeba for a given range of complex structure moduli, we can make the phase diagram for any toric Calabi-Yau 3-fold as finely resolved as desired.

To illustrate this point, we generate using our trained CVAE model $N=389376$ coamoeba for $F_0$
in the same range of complex structure moduli as in \cite{Seong:2023njx}, with component values $c_{ab} \in \{0, \pm 0.5, \pm 1.0, \dots, \pm 5.5 ,\pm 6.0\}$.
Using PCA, we obtain a $m_{\text{PCA}}=2$ phase diagram
shown in \fref{f_fig05}(b),
which is as expected smoother and more finely resolved than the one in \fref{f_fig05}(a).
\\

\begin{figure*}[ht!!]
\begin{center}
\resizebox{0.85\hsize}{!}{
\includegraphics[height=5cm]{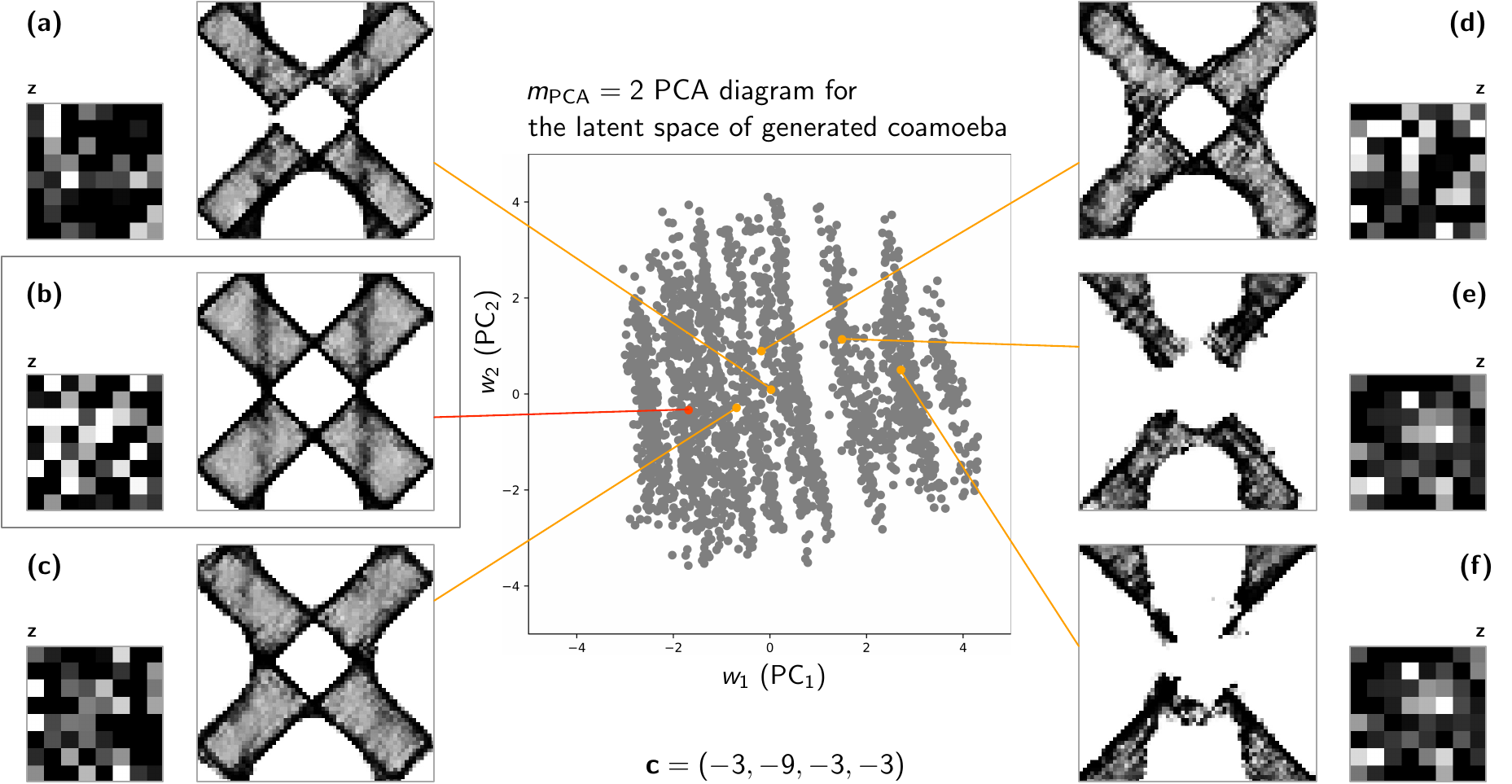} 
}
\caption{
The central plot shows a $m_{\text{PCA}}=2$ dimensional PCA diagram of the $m=64$ dimensional latent space $L$
obtained from training the CVAE model on a training dataset of $N=2304$ coamoeba of $F_0$. 
Based on the trained CVAE model and the reparameterization in \eref{es07a24}, we use the encoder network with optimized parameters $\theta_e^*$
to obtain a collection of $N=2304$ latent vectors $\mathbf{z} = \mu(\theta_e^*) \in \mathbb{R}^{64}$, which we dimensionally reduce using PCA to the $m_{\text{PCA}}=2$ dimensional representation of the latent space shown here in the central plot. 
Every point in the PCA diagram represents a distinct latent vector $\mathbf{z}$, with the one chosen in (b) corresponding to the optimal one given in \eref{es08a02}. 
We can see from the latent space representation that choices for $\mathbf{z}$ that are further away from the optimal one in (b) become less optimal, producing coamoeba with less well-defined boundaries in $T^2$. 
\label{f_fig07}}
 \end{center}
 \end{figure*}

\paragraph{Continuous Paths in the Phase Space.}
A discovery made in \cite{Seong:2023njx} has been that a subset of the components of the complex structure moduli for a toric Calabi-Yau 3-fold determine the toric phases and phase transitions between them.
In the case of $F_0$, 
it is the real component $c_{11}$ 
whose sign determines whether the coamoeba is in phase 1, 2a or 2b, 
as summarized in \eref{es01a21}.
We call $c_{11}$ as an \textit{order parameter} of the phase space for $F_0$.

In terms of such order parameters, and the help of our trained CVA model, 
we are now able to introduce continuous paths in the phase space of a toric Calabi-Yau 3-fold
along which order parameters vary whereas all other components of complex structure moduli stay constant. 
We call these $\textit{phase paths}$, which we can define in the case of $F_0$ as follows, 
\beal{es08a10}
\rho_{(c_{12},c_{21},c_{22})}(t) =
\left(
w_1(\hat{\mathbf{x}})
,
w_2(\hat{\mathbf{x}})
\right)
\Big|_{\mathbf{c}=(t,c_{12},c_{21},c_{22})}
~,~
\nn\\
\eea
where the generated coamoeba $\hat{\mathbf{x}}$ is given by \eref{es07a34},
$w_1$ and $w_2$ are the first two principal components when the generated coamoeba $\hat{\mathbf{x}}$
are dimensionally reduced to a $m_{\text{PCA}}=2$ dimensional phase space,
and $t=c_{11}$ is the continuous order parameter for $F_0$. 
We note here that such phase paths can be easily obtained through the help of the trained CVAE model, which can generate 
coamoeba as instances of the phase path at any arbitrary small intervals of the order parameter $t$. 
\fref{f_fig03}
illustrates the phase paths $\rho_{(-9,-3,-3)}(t)$ and $\rho_{(+6,+3,-3)}(t)$ in the $m_{\text{PCA}}=2$ phase diagram for $F_0$,
with snapshots of coamoeba at various intervals in $t$.

In principle, there are infinitely many phase paths for a given toric Calabi-Yau 3-fold.
All of them connect different toric phases together, as illustrated in \fref{f_fig03} for the case of $F_0$. 
Moreover, they represent paths in the phase space
along which D5- and NS5-branes move continuously in $T^2$, leading from one distinct Type IIB brane configuration to another.
These transitions correspond to Seiberg duality between the realized $4d$ $\mathcal{N}=1$ supersymmetric gauge theories. 
We plan to elaborate in future work more on the different properties of these phase paths.
\\

\paragraph{The Latent Space as a new Moduli Space.}
The CVAE model learns a $m$-dimensional latent space $L$
from the training dataset of coamoeba corresponding to a toric Calabi-Yau 3-fold.
This latent space appears to determine the sharpness of the boundaries of the mirror curve $\Sigma$
projected onto $T^2$ via the coamoeba map.
In turn, this seems to influence the sharpness of the boundaries of the D5-brane and the NS5-brane wrapping $\Sigma$
and their intersections mapped in $T^2$.
The illustration of different coamoeba for $F_0$ in $T^2$ at different values of complex structure moduli $\mathbf{c}$ and latent vector $\mathbf{z} \in \mathbb{R}^{64}$ in \fref{f_fig02} seems to confirm this observation. 
The latent space vector highlighted in \fref{f_fig02}(e) 
corresponds to the one in \eref{es08a02}, which was used to generate coamoeba for $F_0$ in our work. 
This choice appears to produce sharp boundaries for the generated coamoeba on $T^2$.
Additionally, 
it yields
coamoeba that are most consistent with the training dataset used to train the CVAE model as illustrated in \fref{f_fig01}. 

For our example $F_0$, 
based on the reparameterization of the latent vector $\mathbf{z}$ in \eref{es07a24}, 
we can set its components to
\beal{es08a15}
z_k = \mu_k(\theta_e^*)
~,~
\eea
where $k=1, \dots, 64$
and $\theta_e^*$ are the optimized encoder parameters of the trained CVAE model.
Taking the training set of $N=2304$ reference coamoeba for $F_0$
with complex structure moduli components $c_{ab}\in \{0,\pm 3, \pm 6, \pm 9\}$, 
we can use the encoder of the trained CVAE model based on \eref{es07a23}
to obtain $N=2304$ latent space vectors $\mathbf{z}_{s} \in \mathbb{R}^{64}$, with $s=1, \dots, 2305$. 
These in turn using a PCA can be dimensional reduced to a $m_{\text{PCA}}=2$ dimensional \textit{latent space representation}
as illustrated in \fref{f_fig07}. 
In this $m_{\text{PCA}}=2$ dimensional latent space representation, every point represents a choice of latent space vector $\mathbf{z}_s$.
With a fixed choice of complex structure moduli $\mathbf{c}=(-3,-9,-3,-3)$ as shown in \fref{f_fig07}, 
we can see in terms of the $m_{\text{PCA}}=2$ dimensional latent space representation how different choices of $\mathbf{z}_s$ affect the overall shape of the coamoeba. 
The choice depicted in \fref{f_fig03}(b) corresponds to the one in \eref{es08a02}, 
which is the optimal choice used in this work for the trained CVAE model.
Moving away from this choice in the $m_{\text{PCA}}=2$ dimensional latent space representation reveals that the coamoeba with $\mathbf{c}=(-3,-9,-3,-3)$ increasingly exhibits blurred boundaries.
This sometimes leads to instances where the generated coamoeba 
resembles those corresponding to completely different choices of complex structure moduli. 

We recall that the trained CVAE model generates coamoeba $\hat{\mathbf{x}}$
as a probability distribution $q_{\theta_d}(x_{ij}=1|\mathbf{z}, \mathbf{c})$ on $T^2$, as defined in \eref{es07a34}. 
This distribution represents
the probability of the mirror curve occupying the grid point labeled $(i,j)$
in the discretized unit cell of $T^2$, based on \eref{es01a25}.
Consequently, 
the blurred boundaries of the generated coamoeba for certain choices of latent space vectors $\mathbf{z}$
indicate that the probability distribution is spread out rather than sharply peaked. 
This spread reflects an
increased level of uncertainty captured by the latent space vectors $\mathbf{z}$
regarding the actual location of the mirror curve on $T^2$
and the branes in the corresponding Type IIB brane configuration.
\\

We plan to investigate further the physical and geometric significance of latent spaces
in generative AI models such as the CVAE model that we introduce here.
We also plan to apply the capabilities of generative AI models such as the CVAE model 
to construct continuous phase spaces and phase paths for a variety of toric Calabi-Yau 3-folds and corresponding $4d$ $\mathcal{N}=1$ supersymmetric gauge theories, including toric Calabi-Yaus and supersymmetric gauge theories in different dimensions \cite{Franco:2015tya, Franco:2016nwv, Franco:2016qxh, Franco:2016tcm, Franco:2022gvl}. 
We plan to report on these studies in future work. 
\\

\acknowledgments

R.-K. S. would like to thank the Simons Center for Geometry and Physics at Stony Brook University, 
the Kavli Institute for the Physics and Mathematics of the Universe at the University of Tokyo, 
the Interdisciplinary Theoretical and Mathematical Sciences Program at RIKEN, 
the theoretical physics groups at Imperial College London and Queen Mary, University of London, 
the London Institute for Mathematical Sciences, 
as well as the Kavli Institute for Theoretical Sciences at the Chinese Academy of Sciences
and the Yau Mathematical Sciences Center at Tsinghua University
for hospitality during various stages of this work.
He is supported by an Outstanding Young Scientist Grant (RS-2025-00516583) of the National Research Foundation of Korea (NRF).
He is also supported by a Start-up Research Grant for new faculty at UNIST (1.210139.01) and a UNIST AI Incubator Grant (1.240022.01).
He is also partly supported by the BK21 Program (``Next Generation Education Program for Mathematical Sciences'', 4299990414089) funded by the Ministry of Education in Korea and the National Research Foundation of Korea.


\bibliographystyle{jhep}
\bibliography{mybib}

\end{document}

%% file: pref.tex
\newcommand{\be}{\begin{equation}}
\newcommand{\ee}{\end{equation}}
\newcommand{\beq}{\begin{equation}}
\newcommand{\beql}[1]{\begin{equation}\label{#1}}
\newcommand{\eeq}{\end{equation}}
\newcommand{\ba}{\begin{array}}
\newcommand{\ea}{\end{array}}
\newcommand{\bea}{\begin{eqnarray}}
\newcommand{\beal}[1]{\begin{eqnarray}\label{#1}}
\newcommand{\eea}{\end{eqnarray}}
\newcommand{\ben}{\begin{enumerate}}
\newcommand{\een}{\end{enumerate}}
\newcommand{\bean}{\begin{eqnarray*}}
\newcommand{\eean}{\end{eqnarray*}}
\newcommand{\eref}[1]{(\ref{#1})}
\newcommand{\sref}[1]{\S\ref{#1}}
\newcommand{\tref}[1]{Table~\ref{#1}}
\newcommand{\nn}{\nonumber}

\newcommand{\fref}[1]{Figure \ref{#1}}
\newcommand{\btab}[1]{\begin{tabular}{#1}}
\newcommand{\etab}{\end{tabular}}

\def \Black {\pdfsetcolor{0 0 0 1}}

\newcommand{\comment}[1]{}

\newcommand{\qed}{\nobreak \ifvmode \relax \else
      \ifdim\lastskip<1.5em \hskip-\lastskip
      \hskip1.5em plus0em minus0.5em \fi \nobreak
      \vrule height0.75em width0.5em depth0.25em\fi}

\definecolor{darkspringgreen}{rgb}{0.09, 0.45, 0.27}
\definecolor{forestgreen}{rgb}{0.13, 0.55, 0.13}